\documentclass[sigconf]{acmart}

\AtBeginDocument{%
  \providecommand\BibTeX{{%
    \normalfont B\kern-0.5em{\scshape i\kern-0.25em b}\kern-0.8em\TeX}}}

\copyrightyear{2025}
\acmYear{2025}
\setcopyright{cc}
\setcctype{by}

\usepackage{multirow}
\usepackage{soul}
\usepackage[utf8]{inputenc}
\usepackage[linesnumbered,ruled,vlined]{algorithm2e}
\usepackage{tcolorbox}
\usepackage{fontawesome5}
\usepackage{fontspec} 
\usepackage{listings}
\usepackage{tabularx}
\usepackage{graphicx}
\usepackage{booktabs}
\usepackage{xcolor}
\usepackage[normalem]{ulem} 

\IfFontExistsTF{Inter}{
  \newfontfamily\inter[Weight=400,Scale=0.85]{Inter}
}{
  \newcommand{\inter}{\sffamily}
}

\begin{document}


\title{A Text-Native Interface for Generative Video Authoring}


\author{Xingyu Bruce Liu}
\affiliation{%
  \institution{Adobe Research}
  \city{San Francisco}
  \state{CA}
  \country{USA}}
 \email{xingyul@adobe.com}

 \author{Mira Dontcheva}
\affiliation{%
  \institution{Adobe Research}
  \city{Seattle}
  \state{WA}
  \country{USA}}
 \email{mirad@adobe.com}

  \author{Dingzeyu Li}
\affiliation{%
  \institution{Adobe Research}
  \city{Seattle}
  \state{WA}
  \country{USA}}
 \email{dinli@adobe.com}


\definecolor{darkgreen}{rgb}{0,0.5,0}
\definecolor{orange}{rgb}{1,0.5,0}

\definecolor{teal}{rgb}{0,0.5,0.5}
\definecolor{darkpurple}{rgb}{0.5, 0, 0.5}
\definecolor{burntorange}{rgb}{0.8, 0.3, 0}
\definecolor{forestgreen}{rgb}{0.13, 0.55, 0.13}
\definecolor{goldenrod}{rgb}{0.85, 0.65, 0.13}

\definecolor{level-1}{RGB}{215,25,28}
\definecolor{level-2}{RGB}{241, 90, 41}
\definecolor{level-3}{RGB}{251, 196, 64}
\definecolor{level-4}{RGB}{166,217,106}
\definecolor{level-5}{RGB}{26,150,65}

\makeatletter
\expandafter\def\csname emojiicon@clapper-board\endcsname{\faFilm}
\expandafter\def\csname emojiicon@bust-in-silhouette\endcsname{\faUser}
\expandafter\def\csname emojiicon@house\endcsname{\faHome}
\expandafter\def\csname emojiicon@puzzle-piece\endcsname{\faPuzzlePiece}
\expandafter\def\csname emojiicon@framed-picture\endcsname{\faImage}
\expandafter\def\csname emojiicon@artist-palette\endcsname{\faPaintBrush}
\expandafter\def\csname emojiicon@movie-camera\endcsname{\faFilm}
\expandafter\def\csname emojiicon@triangular-ruler\endcsname{\faRuler}
\expandafter\def\csname emojiicon@sparkles\endcsname{\faStar}
\expandafter\def\csname emojiicon@microphone\endcsname{\faMicrophone}
\expandafter\def\csname emojiicon@musical-note\endcsname{\faMusic}
\expandafter\def\csname emojiicon@headphone\endcsname{\faHeadphones}
\newcommand{\emoji}[1]{%
  \ifcsname emojiicon@#1\endcsname
    \csname emojiicon@#1\endcsname
  \else
    \faCircle
  \fi
}
\makeatother



\newcommand{\eg}{\textit{e.g., }}
\newcommand{\ie}{\textit{i.e., }}
\newcommand{\etal}{et al. }

\newcommand {\systemname}{}

\newcommand{\slashmenu}{\icode{/}\xspace}
\newcommand{\shot}{\pillicon{\faFilm}\xspace}
\newcommand{\shotlight}{\pilliconlight{\faFilm}\xspace}
\newcommand{\shotd}{\pillicon{\faImage}\xspace}

\definecolor{sky700}{HTML}{0369a1}
\definecolor{purple700}{HTML}{6b21a8}
\definecolor{rose700}{HTML}{BE123C}

\newcommand{\mention}[1]{{\color{sky700}\inter #1}}
\newcommand{\hashtag}[1]{{\color{purple700}\inter #1}}
\newcommand{\equalsign}{\textcolor{rose700}{\inter =}\xspace}

\newcommand{\newshot}{{\inter \emoji{clapper-board} New Shot}\xspace} 
\newcommand{\mentioncharacter}{{\inter \emoji{bust-in-silhouette} Character}\xspace} 
\newcommand{\mentionscene}{{\inter \emoji{house} Scene}\xspace} 
\newcommand{\mentioningredient}{{\inter \emoji{puzzle-piece} Ingredient}}
\newcommand{\mentionframe}{{\inter \emoji{framed-picture} Frame}\xspace} 
\newcommand{\hashtagstyle}{{\inter \emoji{artist-palette} Style}\xspace} 
\newcommand{\hashtagcamera}{{\inter \emoji{movie-camera} Camera}\xspace} 
\newcommand{\hashtagcomposition}{{\inter \emoji{triangular-ruler} Composition}\xspace}
\newcommand{\hashtagambience}{{\inter \emoji{sparkles} Ambience}\xspace} 
\newcommand{\audiospeech}{{\inter \emoji{microphone} Speech}\xspace}
\newcommand{\audiomusic}{{\inter \emoji{musical-note} Music}\xspace} 
\newcommand{\audiosfx}{{\inter \emoji{headphone} SFX}\xspace}

\newcommand{\mentioncharacterbadge}{\bluebadge {\mentioncharacter} \xspace}
\newcommand{\hashtagstylebadge}{\purplebadge {\hashtagstyle} \xspace}

\newtcbox{\icode}{on line,
  colback=gray!20, colframe=gray!20,
  boxrule=0pt, arc=3pt, boxsep=1pt,
  left=2pt, right=2pt, top=1pt, bottom=1pt}

\newtcbox{\squareicon}{on line,
  colback=gray!20, colframe=gray!20,
  boxrule=0pt, arc=3pt, boxsep=1pt,
  left=1.5pt, right=1.5pt, top=3pt, bottom=3pt}

\newtcbox{\pillicon}{on line,
  colback=gray!20, colframe=gray!20,
  boxrule=0pt, arc=3pt, boxsep=1pt,
  left=4pt, right=4pt, top=1.5pt, bottom=1.5pt}

\newtcbox{\pilliconlight}{on line,
  colback=gray!10, colframe=gray!10,
  boxrule=0pt, arc=3pt, boxsep=1pt,
  left=4pt, right=4pt, top=1.5pt, bottom=1.5pt}

\newtcbox{\handleitem}{on line,
  colback=gray!20, colframe=gray!20,
  boxrule=0pt, arc=3pt, boxsep=1pt,
  left=1.5pt, right=1.5pt, top=1.5pt, bottom=1.5pt, fontupper=\scriptsize}

\newtcbox{\slashitem}{on line,
  colback=white, colframe=white,
  boxrule=0pt, arc=3pt, boxsep=1pt,
  left=4pt, right=4pt, top=1.5pt, bottom=1.5pt}

\newtcbox{\bluebadge}{on line,
  colback=sky700!10, colframe=sky700,
  boxrule=0.5pt, arc=3pt, boxsep=1pt,
  left=2pt, right=2pt, top=0.5pt, bottom=0.5pt, fontupper=\inter\color{sky700}}

\newtcbox{\purplebadge}{on line,
  colback=purple700!10, colframe=purple700,
  boxrule=0.5pt, arc=3pt, boxsep=1pt,
  left=2pt, right=2pt, top=0.5pt, bottom=0.5pt, fontupper=\inter\color{purple700}}

\lstdefinelanguage{json}{
    basicstyle=\ttfamily\footnotesize,
    showstringspaces=false,
    breaklines=true,
    frame=single,
    backgroundcolor=\color{gray!5},
    string=[s]{"}{"},
    morestring=[b]',
    literate=
     *{:}{{{\color{black}:}}}{1}
      {,}{{{\color{black},}}}{1}
      {"shots"}{{{\color{sky700}"shots"}}}{7}
      {"id"}{{{\color{sky700}"id"}}}{4}
      {"context"}{{{\color{sky700}"context"}}}{9}
      {"structuredPrompt"}{{{\color{sky700}"structuredPrompt"}}}{18}
      {"status"}{{{\color{sky700}"status"}}}{8}
      {"tags"}{{{\color{sky700}"tags"}}}{6}
      {"mentions"}{{{\color{sky700}"mentions"}}}{11}
      {"name"}{{{\color{sky700}"name"}}}{6}
      {"tagName"}{{{\color{sky700}"tagName"}}}{9}
      {"hashtags"}{{{\color{sky700}"hashtags"}}}{11}
      {"userPrompt"}{{{\color{sky700}"userPrompt"}}}{13}
      {"target"}{{{\color{sky700}"target"}}}{9}
      {"type"}{{{\color{sky700}"type"}}}{7}
      {"selector"}{{{\color{sky700}"selector"}}}{11}
      {"after"}{{{\color{sky700}"after"}}}{8}
      {"newContent"}{{{\color{sky700}"newContent"}}}{14}
      {"description"}{{{\color{sky700}"description"}}}{14}
}

\begin{abstract}

Everyone can write their stories in freeform text format -- it's something we all learn in school, yet authoring video requires one to learn specialized and complicated tools. In this paper, we introduce \textit{Doki}, a text-native interface for generative video authoring. In Doki, writing text is the primary interaction: within a single document, users define assets, structure scenes, generate shots, refine edits, and add audio. We articulate the design principles of this text-native approach and demonstrate Doki's capabilities through examples. We ran a week-long diary study with 10 participants of varied expertise. Participants produced 46 videos, reporting faster idea‑to‑draft flow, improved coherence through parameterization, and clearer comprehension of narrative structure in the document view; they also surfaced limitations around model predictability, precise control, and temporal expressivity. With Doki, we explore a fundamental shift in generative video interfaces, and demonstrate a powerful and accessible new way to craft visual stories.

\end{abstract}

\ccsdesc[500]{Human-centered computing~Interactive systems and tools}
\ccsdesc[500]{Human-centered computing~Natural language interfaces}

\begin{teaserfigure}
    \centering
  \includegraphics[width=\linewidth]{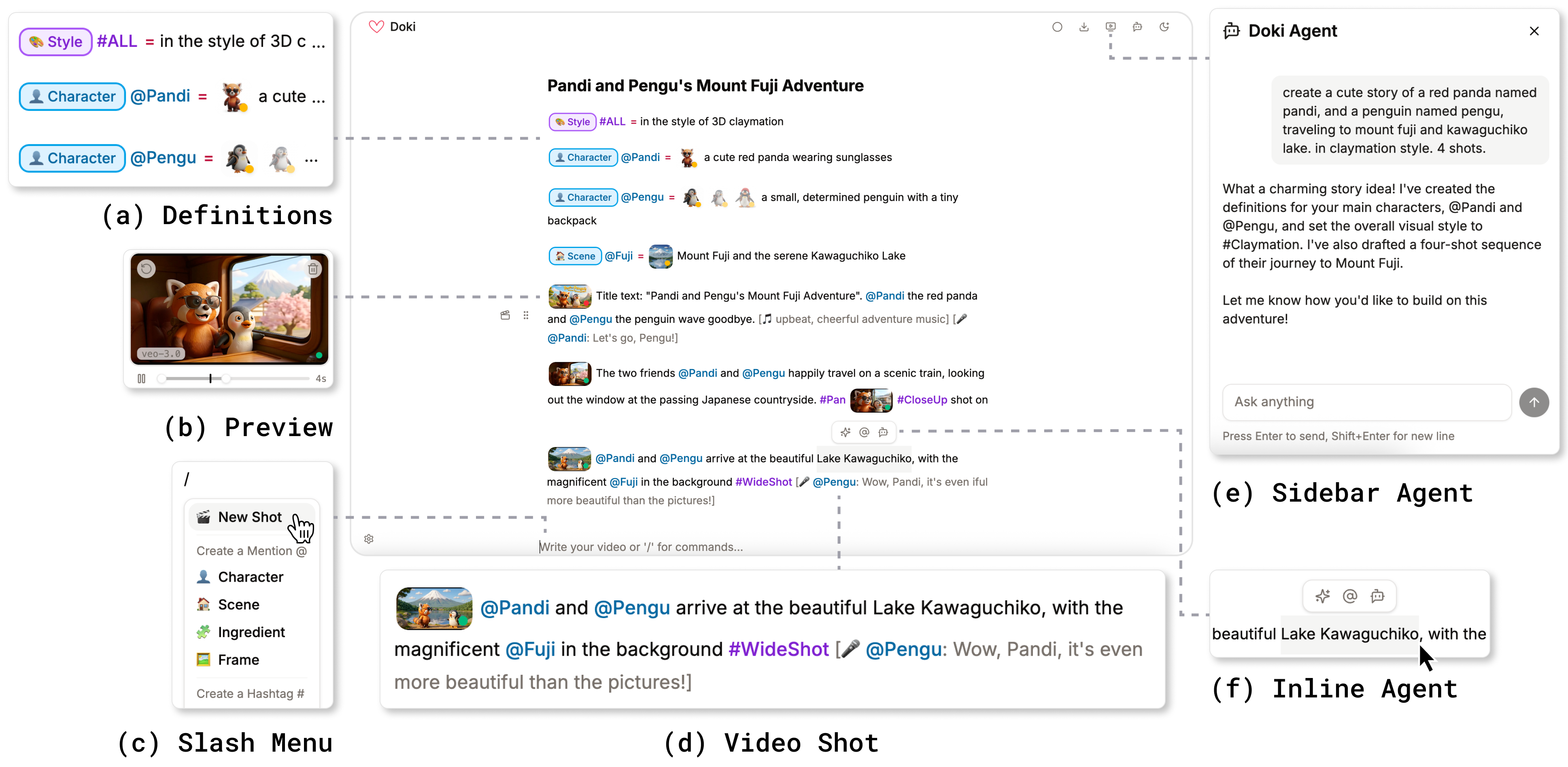}
    \caption{\textit{Doki} is a text-native video authoring interface that creates generative videos within a single document. (a) Users define reusable assets and styles with mentions and hashtags, (b) view and adjust inline previews, (c) access commands through a slash menu, (d) write paragraphs that compile into video shots, and collaborate with both (e) a conversational AI agent and (f) inline agent.}
    \label{fig:teaser}
\end{teaserfigure}

\keywords{Generative AI, Generative Video, AI-Generated Video, Video, Text-to-Video, Video Editing, Creativity Support}



\maketitle

\section{Introduction}

Writing gives form to our imagination and stories.
Children are taught the basics of narrative structure - beginning, middle and end -- as early as first grade. Over time, writing text turns into a core means for communication and self-expression, as we compose emails, articles, or text messages. 
With the rise of video platforms like YouTube and TikTok, video has emerged as a new powerful alternative for communication, expression, and storytelling.
And recent advances in generative AI introduce new possibilities for content creation. As a result, more and more people are authoring video. 

Unfortunately, authoring video remains challenging, because it relies on a very different set of tools than the ones we learn in school. Traditional video non-linear editors emphasize rich interfaces with many capabilities excelling at fine-grained editing tasks at the cost of usability and accessibility. Modern editors, such as Descript~\cite{descript} reduce friction by supporting text-based editing with transcript text, yet such tools only use text for editing speech tracks and adopt the experience of traditional tools with multiple panes and multi-track timelines for authoring visuals. 

Generative AI introduces a paradigm shift in creative tools.
Many call generative models ``the new camera'', because they let people produce high-quality content with a simple text prompt. In this new era, \textbf{text is again a central medium for creative expression}. While most generative video tools~\cite{veo3, sora, runway, pika, kling} focus solely on individual clips and still rely on traditional workflows and timelines for narrative storytelling, we ask:
if audiovisual content can be generated directly from text, \textbf{can video authoring become as natural as editing a document? }

We present \emph{Doki}, a text-native interface for authoring generative videos\footnote{We use the term \textit{generative videos} to denote content produced with generative AI methods. Unlike ``AI-generated videos,'' which can imply AI-led authorship, \textit{generative videos} is neutral about agency and ownership.}. 
In Doki, writing is the primary interaction.
Within a \textbf{single document}, users define assets, structure scenes, generate shots, refine edits, and add audio.
The document serves both as narrative and as an executable script for video production. 
Our design pursues four goals:

(1) \textit{Make text a central medium for authoring.} 
Text is both natural for people and native to AI. It serves as a perfect intermediate common ground between human and AI, as it allows humans to freely edit and quickly review, while enabling AI to generate and suggest using the same medium. Its freeform nature also supports flexible workflows that move fluidly between ideation, generation, and revision.

(2) \textit{Consolidate authoring into a single representation.} 
Current workflows require creators to juggle multiple forms and views of the same video, increasing cognitive load~\cite{chandler1991cognitive}. In Doki, scripts, prompts, visuals, audio, and timelines co-exist in a structured text representation.
While we do not aim for a full replacement for professional toolchains, Doki supports many end-to-end creative tasks.

(3) \textit{Preserve consistency via parameterization.} 
As projects scale, maintaining coherence becomes difficult. 
Doki introduces parametrized definitions to keep characters, styles, and other elements consistent across narratives, and makes it easier to construct complex scenes and develop longer, more cohesive stories.

(4) \textit{Keep interaction simple.} 
Professional tools are powerful but can be overwhelming with complex UI. Doki offers a minimalist design with a lightweight slash menu for creating assets and shots, and inline shot previews for reviewing and editing directly in the document.

We evaluated Doki in a week‑long diary study with 10 participants, spanning filmmakers to first‑time creators. 
Participants submitted 46 videos and on average rated Doki System Usability Scale 81.2 (``Excellent'').
They reported that Doki's text-native structure not only accelerated the transition from ideation to output, but also enhanced their holistic understanding of narrative flow. 
Furthermore, participants found that parameterized definitions provides a reliable structural backbone for cohesive storytelling.
Our findings also reveal that the system's utility scaled distinctly with user expertise. Novices felt newly empowered to author visual stories that were previously out of reach. 
Conversely, domain experts integrated the tool as a complementary engine for rapid ideation and storyboarding, rather than a replacement for high-fidelity production environments.

The study also surfaced insights into human-AI collaboration. 
The document interface encouraged users to heavily delegate production tasks to AI agents. Yet, despite this extensive automation, creators maintained a strong sense of authorship, frequently likening their role to a ``director''. 
Text document serves as a perfect ``common ground'', an intermediate representation that is simultaneously readable and editable by both parties. Humans can freely build the narrative foundation, while AI operations, whether generating drafts or applying edits, remain transparent, legible, and fully revisable by the human.

In summary, we contribute:
\begin{itemize}
\item \textbf{A structured, parametrized text-based representation for videos}. 
We propose a three-layer, hierarchical \textit{``document as video, paragraph as sequence, sentence as shot''} structure, and introduces a definition system (\mention{@mentions} and \hashtag{\#hashtags}) with propagation and context handling to ensure cross-shot consistency. This representation is independent of any particular interface.

\item \textbf{Doki, a minimalistic, text-native interface} that instantiates this representation. 
Users can create videos end-to-end, from ideation to export, within one single document. Doki employs a minimalistic user interface design with a lightweight slash menu, inline previews and AI agents.

\item \textbf{An in-the-wild evaluation of Doki}. Empirical insights into human-AI collaboration dynamics, demonstrating how a text-native interface shifts creator workflows, balances the trade-off between macro-level storytelling and micro-temporal control, and negotiates the boundaries of artistic agency and automation .
\end{itemize}

\section{Related Work}

Our work is situated at the intersection of three research areas: video generation models, novel interfaces for video creation, and HCI research on dynamic documents.

\subsection{Generative Models as Individual Video Generators}

Recent advances in generative AI have produced models capable of synthesizing high-fidelity video clips from text descriptions. 
Commercially available systems like Runway~\cite{runway}, Pika~\cite{pika}, Kling~\cite{kling} and Google's Veo~\cite{veo3} demonstrate increasingly powerful capabilities. 
The primary interaction model for these tools is to write a text prompt to generate a single, short, fixed-length video. 
These prompts often require a structured format, specifying elements like the subject, action, context, and desired visual style~\cite{veo3}. 
More advanced features allow for greater control, such as using reference images to guide generation or maintain character consistency~\cite{runway}, and using start and end frames to define transitions~\cite{lumalabs, veo3, pika}.

However, the focus of these models remains on the generation of \textit{individual, short clips}, typically lasting only a few seconds, \eg 8 seconds for Veo 3. 
There is limited to no support for structuring these shots into a longer, coherent narrative. 
Users must manually assemble the generated clips in a separate editing environment. 
Doki builds upon the capabilities of such models. 
It treats them as the supporting engine while contributing a higher-level authoring interface that shifts the interaction from generating isolated clips to authoring a complete video story.


\subsection{Novel Video Authoring Interfaces}

\begin{figure}[t]
\centering
\includegraphics[width=\linewidth]{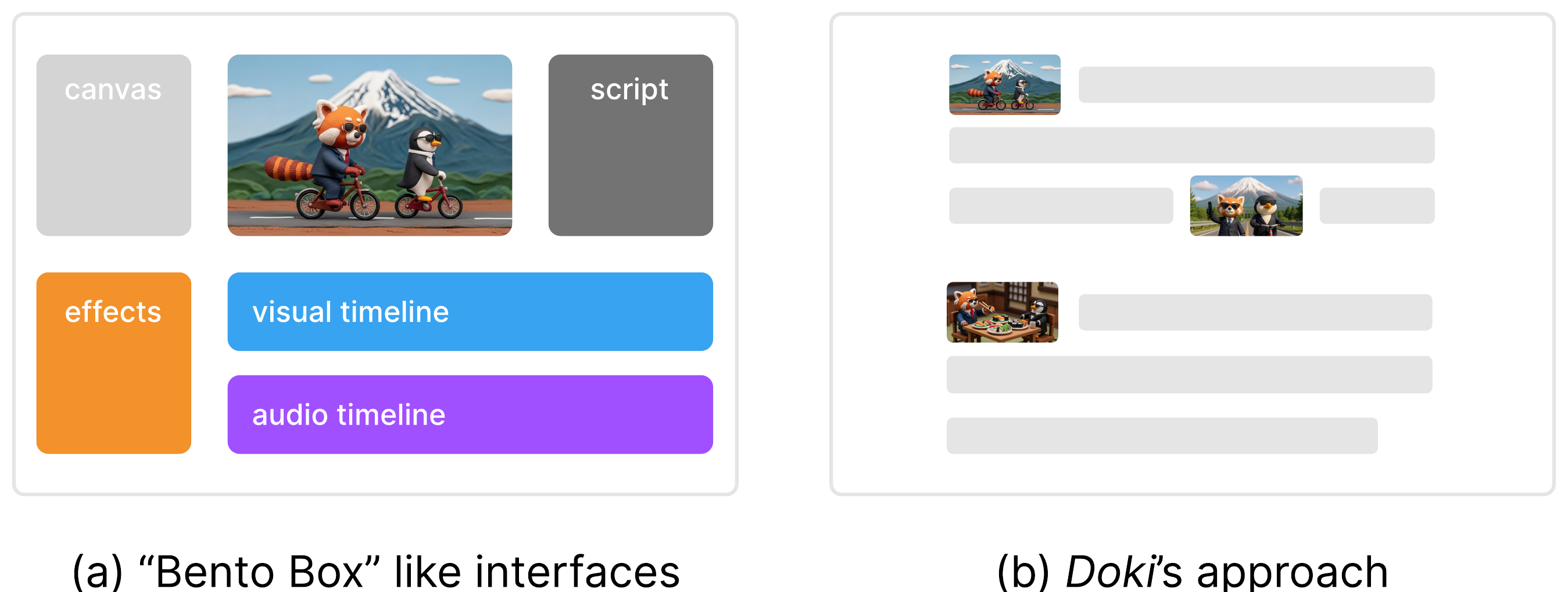}
\caption{A comparison of interface paradigms. (a) ``Bento Box'' style interfaces distribute authoring across multiple, separate representations. (b) Doki's approach uses a text-native canonical representation where the document serves as the primary interface.}
\label{fig:bentobox}
\end{figure}

A growing body of HCI research explores AI-powered interfaces for video authoring beyond traditional non-linear editors. 
Videorigami, for example, organizes authoring around multiple \emph{compositional structures} (\eg canvas, script, storyboard, timeline) and synchronizes them to support human–AI co-creation~\cite{cao2025compositional}. 
LAVE augments editing with AI agent and language annotations of footage~\cite{wang2024lave}.
ExpressEdit lets creators issue multimodal edit intents by combining natural language and on-frame sketching~\cite{tilekbay2024expressedit}.
While these systems explore new interaction modalities, they typically distribute authoring across multiple panes -- what we refer to as a ``bento box'' interface (\autoref{fig:bentobox}a) -- which can incur split-attention costs when creators must reconcile multiple views and formats~\cite{chandler1992splitattention}. 

Doki embraces a contrasting design philosophy aimed at consolidating all authoring steps into a single representation. 

\paragraph{Subtractive vs. additive workflows.}
Due to the nature of video capture prior to generative video models, most existing tools adopt \emph{subtractive} workflows: creators begin with pre-recorded footage (or a library) and remove, trim, and re-order material. Examples include QuickCut~\cite{truong2016quickcut}; transcript-aligned manipulation of talking-head footage~\cite{fried2019talkinghead}; and multimodal edit commands that operate on existing clips~\cite{tilekbay2024expressedit}. 
Some systems explore \emph{additive} workflows where the tool synthesizes assets from text, such as Doc2Video’s document-to-talking-head prototyping~\cite{chi2022doc2video}, but their system is limited to specific shot types rather than authoring complete, multi-shot stories. 
Doki takes an additive stance end-to-end: text is the substrate for \emph{generating} shots and assets, not only selecting or trimming them.

\paragraph{Transcript-based editing.}
Transcript-centered tools let users edit a video by editing its text transcript, with changes reflected on the timeline~\cite{descript, rubin2013content, huh2023avscript, chi2022doc2video}. This interaction is especially effective for dialogue-driven media such as interviews and podcasts, where transcript and timeline are tightly aligned. 
Other systems use text to guide the assembly of existing footage: Write-A-Video~\cite{wang2019writeavideo} selects shots that semantically match an edited paragraph, while Crosscast and Crosspower augment audio with retrieved images~\cite{xia2020crosscast, xia2020crosspower}.
Doki complements these approaches by extending the role of text from mirroring transcript to \emph{generating the full video} including audio tracks and video frames. Rather than treating text as just a proxy to transcript/speech tracks, Doki uses text to create and manage the visuals—and associated audio—end-to-end.

\subsection{Rich-Text Editing and Dynamic Documents}  

Doki's interface design is inspired by research on dynamic documents~\cite{soulver, potluck, embark, promptcanvas, victorExplorable}. Unlike applications that enforce predefined feature sets and rigid task boundaries, dynamic documents are designed as permissive mediums. Much like writing on paper, they do not impose a strict schema, allowing users to capture ideas in freeform ways. This flexibility supports a workflow of ``gradual enrichment''~\cite{potluck}, where users can begin with unstructured text and progressively add structure and computational behavior as their needs evolve. Research prototypes such as Potluck~\cite{potluck} and Embark~\cite{embark} demonstrate this principle by allowing notes to evolve into personal software or interactive trip plans.  

Commercial tools have adopted related approaches. Notion~\cite{notion} provides a block-based workspace where users can draft in natural text and later enrich content. OpenAI's Canvas mode~\cite{openaiCanvas} illustrates a similar trajectory. It extends chat-based interaction into an editable document environment where users can directly edit and request inline assistance. 

Doki builds on these precedents but extends them into the video authoring domain. It treats video creation as the authoring of dynamic documents, that can be read as narratives and executed as a script, supporting a more flexible authoring process than existing video tools.

\section{How People Create Generative Videos Today}

We examined current end-to-end workflows for creating generative videos and describe motivating scenarios for Doki. Our analysis drew on public creators' generative AI video tutorials on YouTube and X~\cite{dankieft_channel, pjacefilms_channel, aivideoschool_channel, kevinstratvert_channel}, and on observations of creators within our organization. 

\subsection{Motivating Scenarios}

\subsubsection{Creator A: Asset-First}
Creator A anchored storytelling in asset construction. She typically spent 3-5 hours creating characters with text-to-image models and used the results as reference frames for consistency across a project. 
Yet, consistency remained elusive even with reference images. As she explained, \textit{``I actually had two different octopus character reference frames that I used in GPT Image when I wanted more or less exhausted.''} 
Because outputs often diverged, she repeatedly checked generation outputs and manually adjusted assets. 
After preparing characters and environments, she created storyboards, generated short clips, and finally imported everything into a separate timeline editor to add voice and music. 
Her workflow required constant switching between asset creation tools, storyboard generators, video models, and editing software. 
Each step relied on a different view or interface, forcing repeated translation between formats.

\subsubsection{Creator B: Script-First}
Creator B started with a complete script -- often drafted with a large language model -- and decomposed it into scenes. Each scene was expanded into a detailed prompt specifying setting, character, tone, and cinematographic parameters. Because current video models lacked context awareness, every shot required re-describing all elements to preserve consistency. He explained, \textit{``I would re-describe the setting, the character, and the tone every time.''} 
This led to long and repetitive prompts, rendering small edits nearly impossible. Once clips were generated, they were reviewed and assembled in a video editor. 
This process prioritized prompt construction over narrative development.

\subsubsection{Creator C: Iterative-Exploratory}
Creator C worked in a highly exploratory style. Rather than preparing a script or asset library, he generated clips rapidly and used them to guide the next step. This improvisational process enabled serendipitous discoveries but became fragile during revision. For example, shifting the story setting from Tokyo to Barcelona required revisiting every prompt and regenerating all clips. 
As with others, he then manually reassembles on a timeline, further slowing iteration.

\subsubsection{Recurring Challenges in Generative Video Authoring}

Although creators approached generative video authoring very differently, many encountered the same set of challenges:

\noindent\textbf{C1. Fragmented Tools and Formats:}
Even for a simple short video, many creators end up using three to five different tools — for example, a text editor for scripts, an image generator for visuals, a video model for clips, audio tools for narration or music, and a non-linear editor to pull it all together. The core elements of a project — scripts, prompts, images, audio, and edits — often live in separate formats across these tools.

This fragmentation can lead to constant context-switching, breaking creative flow and making it harder to maintain momentum. It also makes it difficult to get a clear overview of the project or understand how one change affects others. This raises the barrier to entry and slows iteration.


\noindent\textbf{C2. Prompt Over Storytelling:}
Creators often spend disproportionate time crafting and tweaking prompts, which shifts focus away from narrative development and toward prompt engineering. For example, a 30-shot video might require 30 separate, verbose prompts -- often managed manually in a list. Building up a complex prompt is itself difficult, and even once crafted, it is hard to control or manage across shots.

As a result, the workflow becomes less about shaping a story and more about wrestling with prompts, leaving creators with less bandwidth for the higher-level creative choices that should drive their work.

\noindent\textbf{C3. Consistency and Coherence:}
Maintaining visual and narrative consistency remains difficult. While recent models such as Nano Banana~\cite{google2025gemini2_5flashimage} and Flux Kontext~\cite{bfl2025flux1kontext} can generate consistent images from references, we argue that true coherence is as much an authoring and representation problem as it is a modeling problem. Conditioning on a few reference images does not provide the structural backbone needed to preserve assets, styles, and relationships across an extended narrative. Without a representation that defines elements, propagates them through time, and manages context automatically, consistency will inevitably drift as projects grow.


\subsection{Design Principles}

Our analysis of existing workflows reveals a clear opportunity for a new authoring paradigm for generative videos. We propose four core design principles that guide the creation of Doki:

\noindent\textbf{D1. Make text the central medium for authoring.}
C2 shows how prompt engineering can pull attention away from storytelling. Doki addresses this by making text the primary medium for authoring: prompts are embedded directly into the narrative, so the creative process feels more like writing a story than managing a command list.

Text has several advantages as a medium: it is familiar and accessible for humans, the native input modality for generative models, and inherently flexible and free-form. Because both humans and AI operate in the same medium, revisions and suggestions are immediately understandable: for example, AI-generated revisions are immediately understandable and can be refined in-place by the user, enabling fast and straightforward iteration.

\noindent\textbf{D2. Consolidate video authoring in a single representation.} 
In traditional workflows, creators switch between multiple views of the same project, which fragments the process and increases cognitive load (C1).
Doki instead unifies scripts, prompts, visuals, audio, and timelines into a single canonical text representation. This representation functions both as the editable document that authors work with and as the underlying schema for the video itself.
While not a full replacement for professional toolchains, this unified form enables many end-to-end creative tasks without leaving the document.

\noindent\textbf{D3. Preserve consistency through parameterization.} 
As projects scale, maintaining coherence across characters, styles, and scenes becomes increasingly difficult (C2, C3).
Doki addresses this with a system of parameterized definitions that ensures elements remain consistent while still allowing flexible variation. 
In addition, Doki automatically retrieves relevant references from earlier scenes so that continuity is preserved across the narrative.
Together, these mechanisms make it easier to construct complex scenes, reuse components, and extend projects into longer, more cohesive stories.

\noindent\textbf{D4. Keep interaction simple.} 
Traditional non-linear editors are powerful but can be overwhelming (C1). 
Doki offers a simple editing environment with a lightweight slash menu for generating shots, defining assets, and inserting audio. Authors can write, generate, and edit within the paragraph, while using inline previews to review and trim.

\begin{figure*}[t]
\centering
\includegraphics[width=\linewidth]{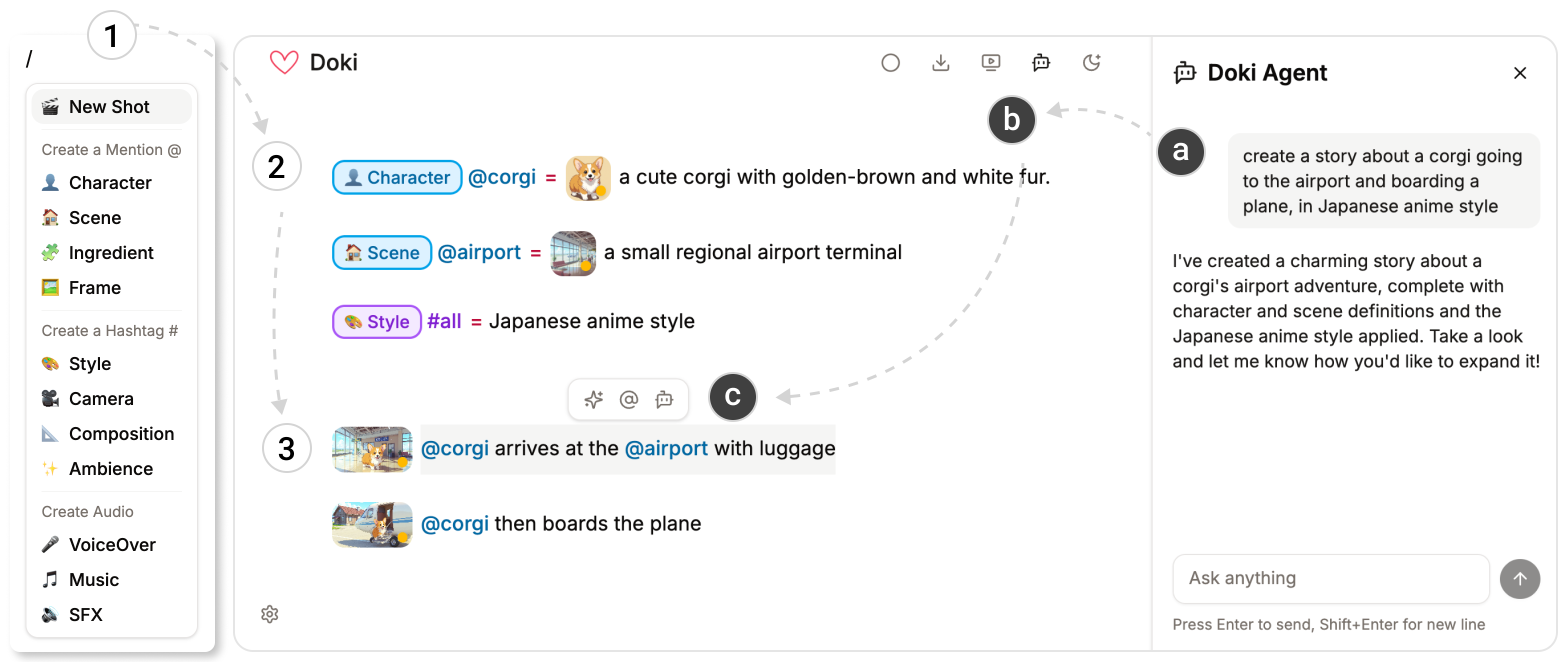}
\caption{Two basic example workflows in Doki. Alice: (1) define assets and shots with slash commands $\rightarrow$ (2) write story and generate previews $\rightarrow$ (3) create video shots; Bob: (a) prompt the sidebar agent for a draft $\rightarrow$ (b) review the AI-generated draft $\rightarrow$ (c) refine with inline AI agent.}
\label{fig:workflow}
\end{figure*}

\section{Doki}
In this section, we first show example workflows of using Doki, then describe the structured text representation. Finally, we walk through each component of the system.

\subsection{Example Workflows}

Two creators produce the same short animation in different ways (\autoref{fig:workflow}).
The story follows a cute little corgi embarking on an adventure.

\subsubsection{Alice: Writing a Video from Scratch}

Alice prefers a hands-on, deliberate process. 
Starting with a blank text editor, she first defines the main assets for her story using slash commands. She types \slashmenu and selects \mentioncharacter from the menu to create a definition block. Here, she names the character \mention{@corgi} and provides a textual description: {\inter “a cute corgi with golden-brown and white fur”}. Similarly, she defines the setting with \slashmenu $\rightarrow$ \mentionscene, naming it \mention{@airport} and describing it as {\inter “a small regional airport terminal”}. To ensure a consistent visual aesthetic, she creates a global style definition using \slashmenu $\rightarrow$ \hashtagstyle, naming it \hashtag{\#all} and adding her description. 
For each definition, Alice generates a preview image. This allows her to check the visual interpretation and use the image as a reference to maintain consistency.

Next, Alice writes the storyline directly in the document. She structures it in paragraphs each corresponding to a video clip. For each clip, she creates a shot by selecting new shot from the slash menu (\slashmenu $\rightarrow$ \newshot). She writes the first shot: {\inter \mention{@corgi} arrives at the \mention{@airport} with luggage}, and the second shot: {\inter \mention{@corgi} then boards the plane}. 
She then generates an image preview for each shot, following by the video.
Once all shots are ready, she can export the video or view it in the built-in player.

\subsubsection{Bob: Directing AI agents}

In contrast, Bob prefers to leverage an AI agent workflow. 
Instead of starting from scratch, Bob uses the sidebar AI agent to generate the initial script.
He provides a one-liner prompt: {\inter “a story about a corgi going to the airport and boarding a plane, in Japanese anime style”}.
The agent generates a complete document structure, including definitions for the character, scene, and style, along with two paragraphs of text for the initial shots.

To refine the agent's output, Bob can manually 
edit the document and the description of \mention{@corgi}. Or, he can invoke the inline AI agent.  For example, he selects the first paragraph and uses the following prompt: {\inter “add some background music here”}. The AI agent understands the context and appends this music description to the paragraph. 

This approach also extends to revisions. Alice can manually change the character from a \mention{@corgi} to a \mention{@cat}, and Doki will regenerate the shots based on those specific changes. And Bob can use the sidebar agent for an agentic revision. He can prompt it to {\inter “adapt the story to happen in a New York City with a cat as the main character”}. The agent then revises the entire document, updating all definitions and shots to fit the new theme. 

The examples above only demonstrate the basic usage of Doki. In real-world use, participants employ more intricate and blended workflows, which we examine in \autoref{sec:deployment}.

\subsection{A Structured Text Representation for Generative Videos}
\label{sec:structure}


\paragraph{Document as Video, Paragraph as Sequence, Sentence as Shot}

At its core, Doki adopts a three-level hierarchical structure that draws inspiration from video production: the document as a whole maps to a video, paragraphs map to sequences, and sentences map to individual shots. This mirrors both the logic of filmmaking and the way humans naturally structure writings.

At the highest level, Doki conceptualizes the entire document as a video project. The arrangement of text determines the structure of the resulting video. 

Within this framework, each paragraph corresponds to a sequence. Just as sequences in film bring together multiple shots to form a coherent scene, paragraphs gather sentences around a continuous narrative. 
A new paragraph often marks a shift in focus, such as transitioning from one topic to another, which parallels the cinematic logic of moving from one scene to the next.

At the most granular level, sentences maps to shots. In filmmaking, a shot is 
a single, uninterrupted take. In Doki, authors insert a shot preview element into the document to begin a new shot. All the sentences that follow this marker, up to the next shot or paragraph break, are treated as the textual description of that shot.

This three-level hierarchy resonates with how writers already think and compose: when beginning a new topic, they start a new paragraph; when articulating a small unit of idea, they write a sentence. Doki aligns these practices with the structure of video production.

\paragraph{Parametrization and Context}
Beyond the basic document structure, Doki enriches this text representation through parametrization and context handling. 
We introduce a definition system that allows authors to define and reuse key elements consistently across the document. It also supports automatic reference resolution, ensuring continuity across sequences and shots. 
Implementation details of these mechanisms are elaborated in the following sections.

\subsection{Doki System}
The Doki interface is intentionally minimal (\autoref{fig:teaser}). 
Upon opening, users see a blank document with only a few controls to learn. In the top right corner, the interface provides a status indicator, download, video player, AI agent, and a dark mode toggle. In the bottom left corner, a settings button is available. 
The central interaction space is the document itself.

\subsubsection{Slash Menu}

When users type a slash \slashmenu (\autoref{fig:workflow}), a menu appears that provides quick access to key creation tools. From this menu, they can start a new video shot, create reusable definitions like characters and scenes, or add audio to their project.

\subsubsection{Shots}
In Doki, \textit{shot} is the fundamental unit of a visual story. Each shot is embedded directly in the text as an inline element, tightly woven into the flow of the narrative. This design allows writing and visuals to coexist in the same space.

\begin{figure}[t]
\centering
\includegraphics[width=\linewidth]{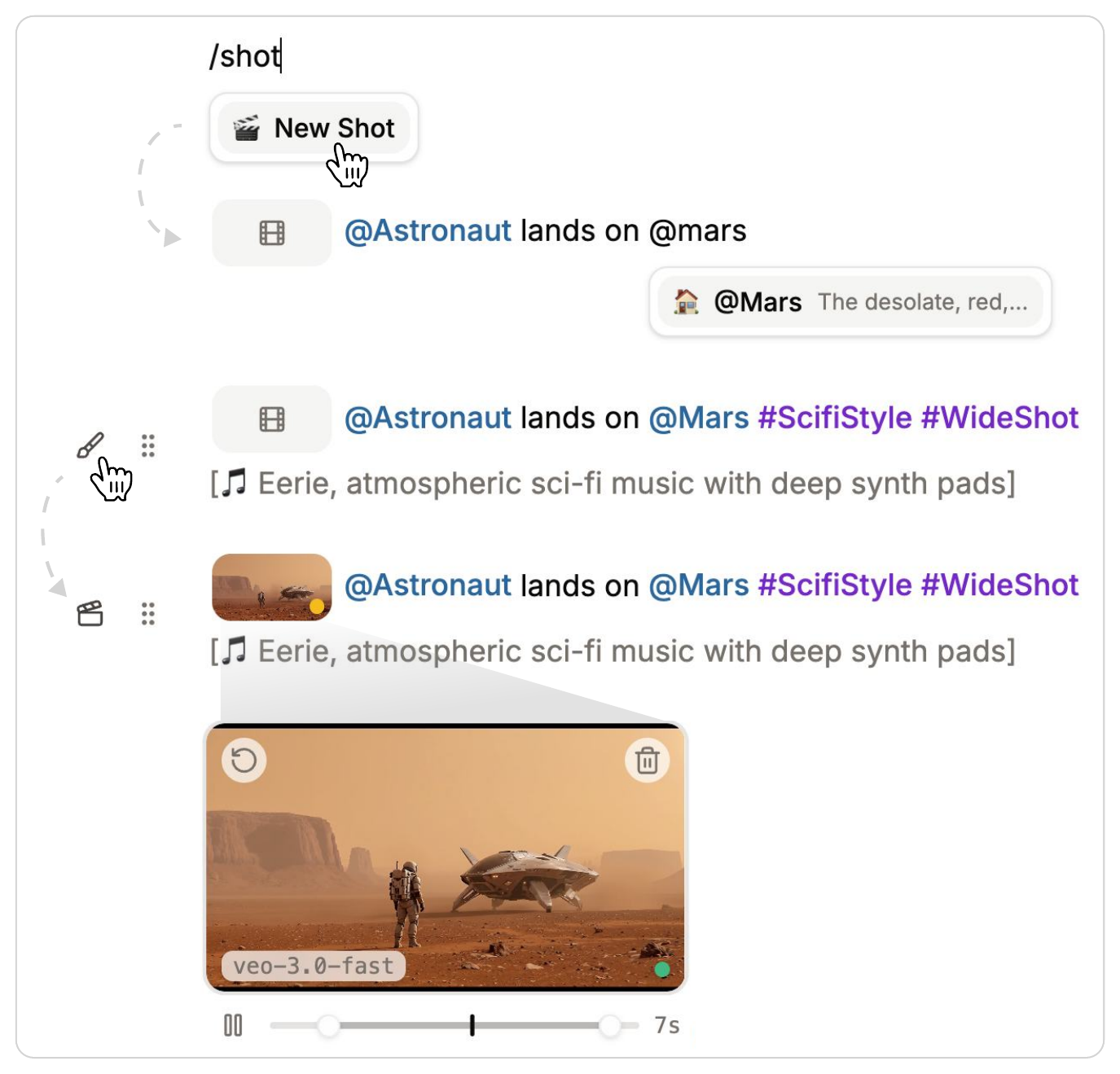}
\caption{Creating a shot in Doki. A new shot is inserted inline with a slash command, and the description that follows serves as its prompt. The system first generates a preview image, which can then be turned into a video clip. Users can click to expand them for playback and additional controls.}
\label{fig:shot}
\end{figure}

\paragraph{Creating a Shot.} To create a shot (\autoref{fig:shot}), a user types \slashmenu $\rightarrow$ \newshot, which inserts a shot preview node \shot. The description that follows the node serves as the raw prompt.

\begin{figure}[t]
\centering
\includegraphics[width=\linewidth]{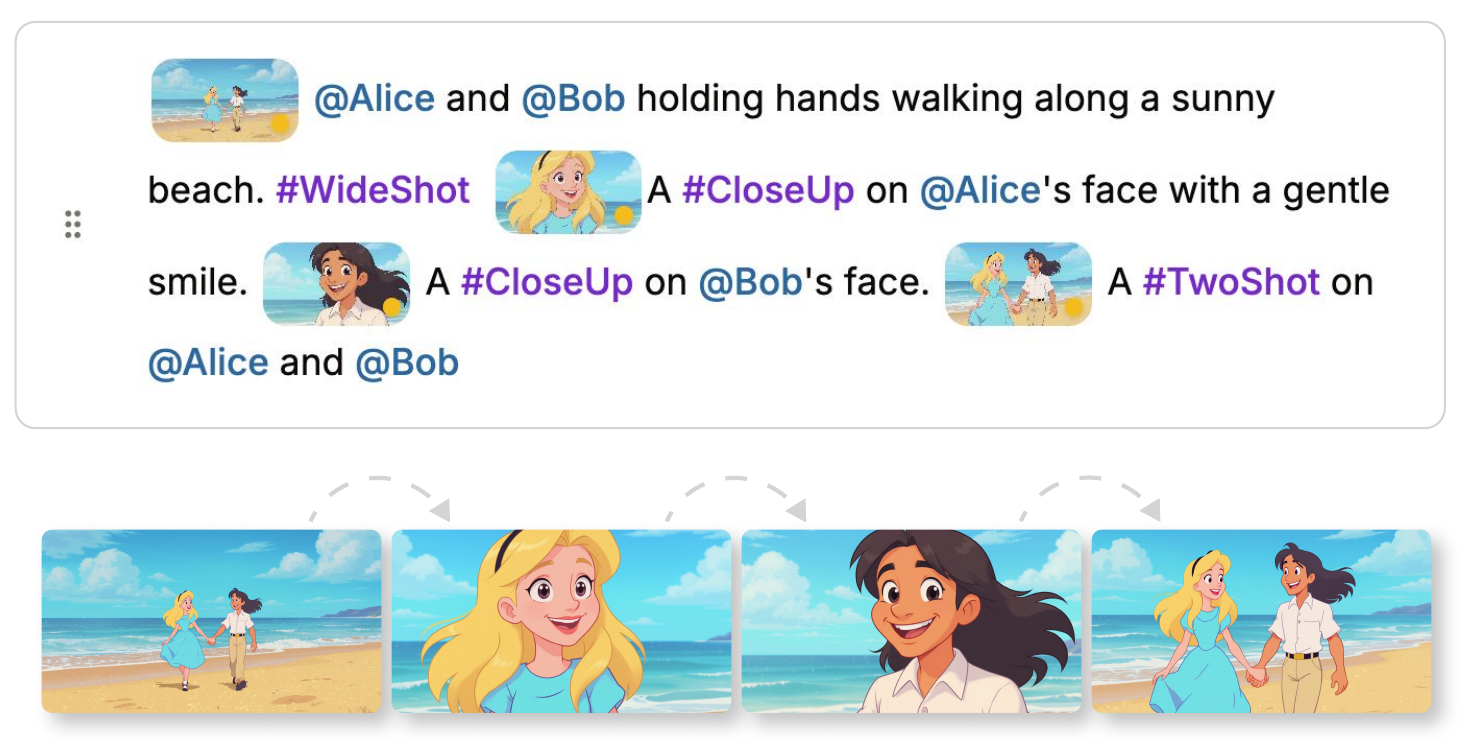}
\caption{Writing consecutive shots within a single paragraph. Later shots inherit context from earlier ones, enabling continuity across a sequence. We achieve strong consistency between shots without repetitive context description in the Doki document.}
\label{fig:shot_sequence}
\end{figure}

\paragraph{Shot Sequences.} As we described in \autoref{sec:structure}, paragraphs in Doki function as analogues to sequences. Users may insert multiple shots within a single paragraph (\autoref{fig:shot_sequence}). In this case, later shots use the image generated by the preceding shot as contextual input for the model. This supports continuity across a sequence. For example, writing:
\begin{quote}
{\inter \shot \mention{@Pandi} riding a bike. \shot \hashtag{\#CloseUp} on \mention{@Pandi}} 
\end{quote}
produces two connected shots. The second builds on the first by applying a close-up shot on the character in the same scene.

\paragraph{Shot Variants.} Users can generate multiple variants (\autoref{fig:variants}) of a shot by right-clicking and selecting {\inter Add Variants}. For example:
\begin{quote}
{\inter \shot \shotlight \shotlight \mention{@Pandi} riding a bike. } 
\end{quote}

This action creates three alternative previews using the same prompt. The first is selected by default, while the others appear semi-transparent as backups. Users can drag any variant to the front, establishing it as the active version.

\paragraph{Generating Image and Video.} 
Each shot follows a staged generation pipeline: text $\rightarrow$ image $\rightarrow$ video. The system first uses the shot’s description to generate a preview image, which then will be used as the first frame for video generation. The inline shot preview node will be updated accordingly as this progress proceeds.
This staged pipeline offers authors control and aligns with existing creative workflows.
Our formative studies revealed that creators typically prefer generating an image before producing video, as it provides an opportunity to preview visual direction before committing to a full video. Cost differences reinforce this preference: generating video costs \$3.20 per clip (Veo 3), compared to \$0.04 per image generation (Imagen 4). 
For shots in the same paragraph, Doki processes each shot in order, using prior images as contextual references for subsequent ones (details in \autoref{sec:handles}).


\paragraph{Shot Status.} Each shot displays its current status through a small circular indicator in the bottom-right corner. The indicator communicates whether a shot is generating an image (breathing yellow), image ready (static yellow), generating a video (breathing green), video ready (static green), or outdated (red). A shot becomes outdated when its prompt or any referenced definition changes.


\paragraph{Expanded Shot.} While shots are compact inline elements by default, they can be expanded with a click. This expanded view (\autoref{fig:shot}) provides additional controls, allowing users to trim videos, delete, or regenerate. Model information is also displayed.


\subsubsection{Parametrized Definitions}
Definitions are reusable parametrized building blocks in Doki. They allow creators to specify characters, scenes, styles, and other elements once and reference them throughout the document. 

\paragraph{Types of Definitions.}
Doki supports two categories of definitions: \mention{@Mentions} and \hashtag{\#Hashtags}. Mentions define elements of the story, while hashtags specify qualities that shape how the story appears. Mentions act like nouns, while hashtags act like adjectives.

\mention{@Mentions} include:
\begin{itemize}
\item \mentioncharacter — a character in the story (\eg \mention{@Pandi}, a little red panda wearing a suit and sunglasses).
\item \mentionscene — a location or setting (\eg \mention{@subway}, a crowded subway cabin in Tokyo).
\item \mentioningredient — an object or prop (\eg \mention{@sushiBox}, a wooden bento box filled with sushi).
\item \mentionframe — a reference frame (\eg \mention{@PandiSubway}, a frame of \mention{@Pandi} in the \mention{@subway}).
\end{itemize}

\hashtag{\#Hashtags} include:
\begin{itemize}
\item \hashtagstyle — the overall visual aesthetic (\eg \hashtag{\#filmnoir}, \hashtag{\#claymation}, \hashtag{\#anime}).
\item \hashtagcamera — cinematic movements (\eg \hashtag{\#closeUp} on \mention{@Pandi}, \hashtag{\#Pan} across \mention{@TokyoTower}).
\item \hashtagcomposition — spatial arrangement or framing (\eg \hashtag{\#twoShot}, \hashtag{\#wideShot}).
\item \hashtagambience — mood, color, or lighting (\eg \hashtag{\#sunsetGlow}, \hashtag{\#neonLights}, \hashtag{\#misty}).
\end{itemize}

Doki also provides a built-in cinematography library with professional terms such as \hashtag{\#CloseUp}, \hashtag{\#Pan}, and \hashtag{\#Dolly}, making it easy to apply established film techniques.

\begin{figure}[t]
\centering
\includegraphics[width=\linewidth]{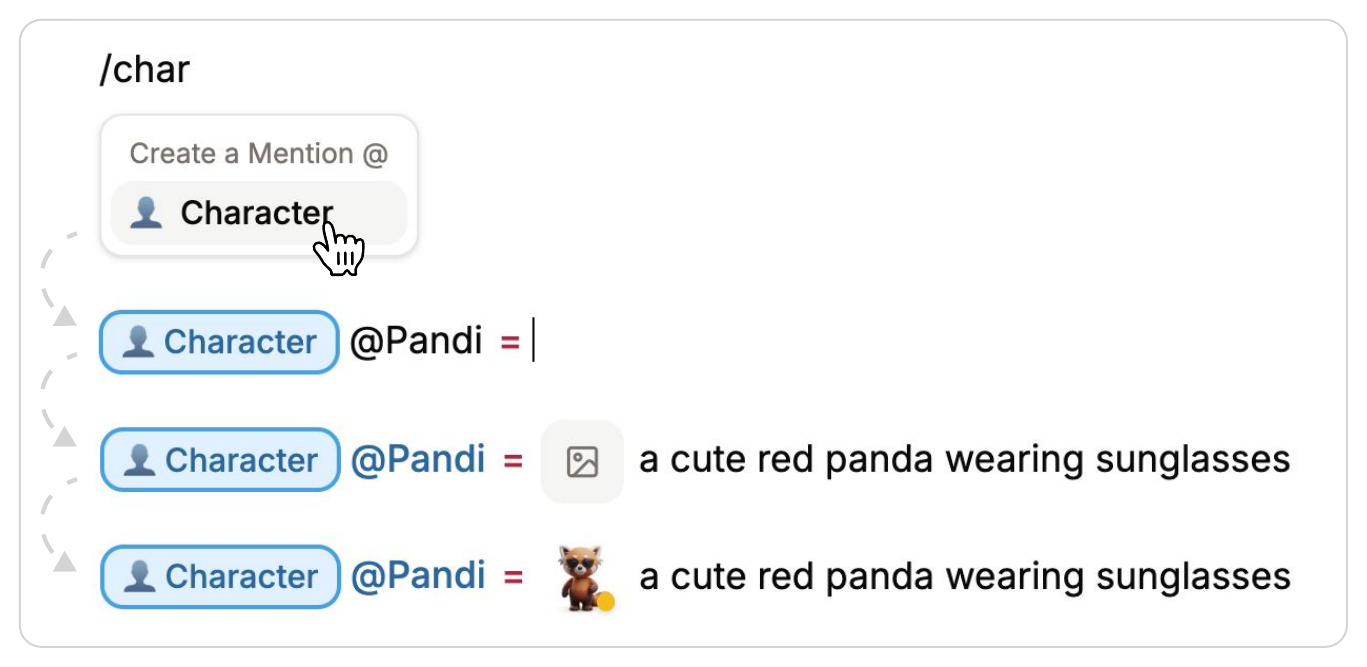}
\caption{Creating definitions in Doki. Users open the command menu with a \slashmenu, select a type, and provide a name and description. Optionally they can add a visual definition for even better consistency.}
\label{fig:definition}
\end{figure}

\paragraph{Creating Definitions.}
To create a new definition, users type the slash \slashmenu to reveal the menu, then select the definition type. This inserts a tag indicating the definition type, followed by the appropriate prefix symbol (either \mention{@} for mentions or \hashtag{\#} for hashtags). 
Immediately after the prefix, users first provide a name for the definition, then enter an equals sign \equalsign and a natural language description (\autoref{fig:definition}).
For example:

\begin{quote}
{\inter \mentioncharacterbadge\xspace \mention{@Pandi} \equalsign a little red panda}
\end{quote}


Definitions can be combined and nested. For instance, one character definition may include references to other characters, props, or styles. This allows users to incrementally build more complex concepts from simpler ones, while maintaining reusability across the document.

\begin{figure}[t]
\centering
\includegraphics[width=\linewidth]{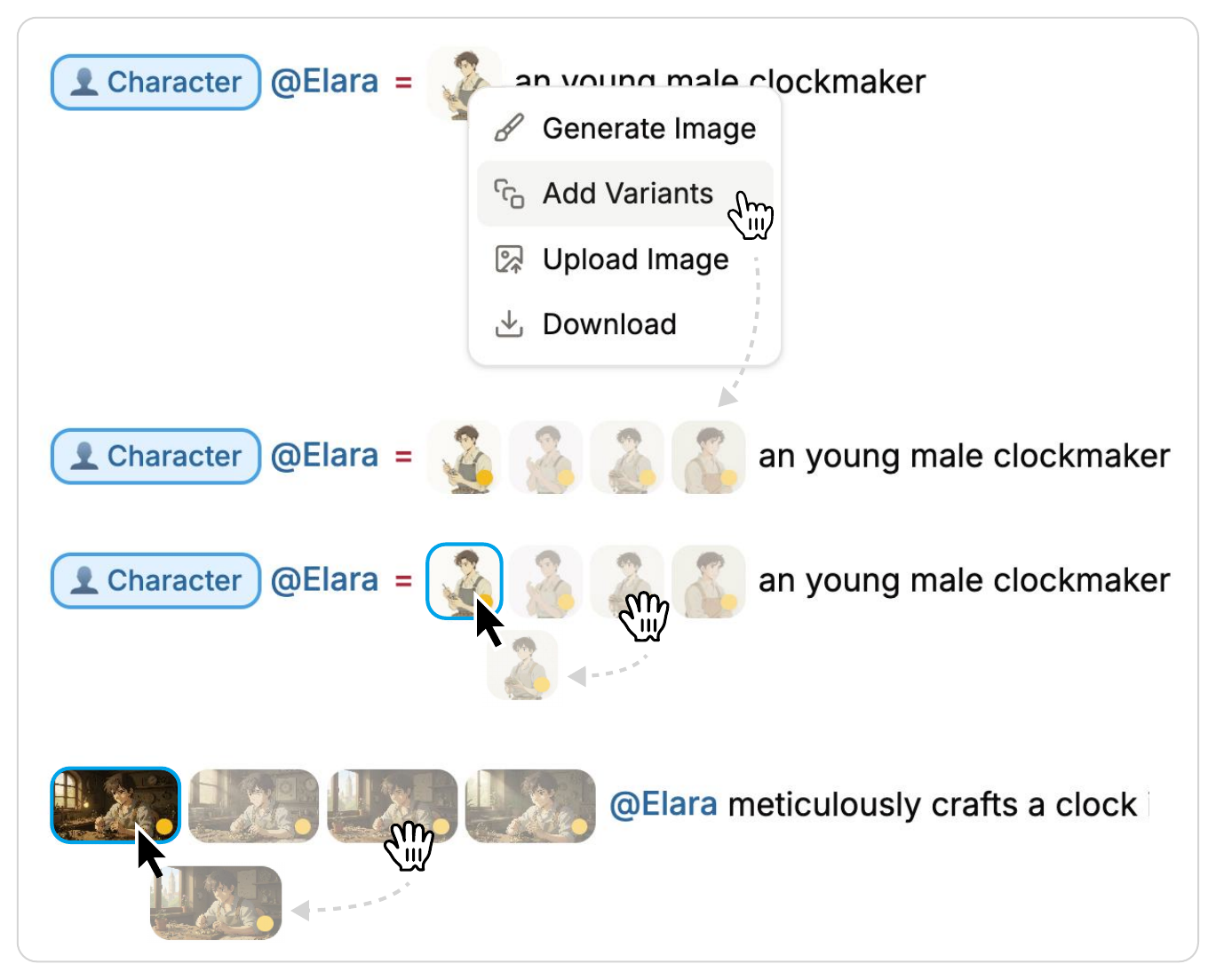}
\caption{Adding variants to a definition and shot. The user selects {\inter Add Variants} from the context menu by right-clicking on the image. Three variations will be added inline. The user can then select one variant by moving it to the first position.}
\label{fig:variants}
\end{figure}

\paragraph{Visual Definitions.}
In addition to textual descriptions, users may attach a visual preview to any definition. To do so, insert a shot using the slash command \slashmenu $\rightarrow$ \newshot immediately after the equals sign (\autoref{fig:definition}):

\begin{quote}
{\inter \mentioncharacterbadge\xspace \mention{@Pandi} \equalsign \squareicon{\faImage} a little red panda}
\end{quote}

If a visual is added, Doki uses it as the primary reference for all future generations involving that asset. If no visual is present, the system falls back to the text description. 
Similar to shots, visual definitions can have multiple variants (\autoref{fig:variants}).
In addition, Doki also supports user-uploaded images for definitions. A user can upload a custom image by selecting {\inter Upload Image} in the right-click context menu.
The system automatically analyzes the image and generates a corresponding textual description using Gemini 2.5 Flash.

\paragraph{Referencing Definitions}
Once a definition is created, it can be reused throughout the document. 
To reference a definition, the author types either \mention{@} or \hashtag{\#} followed by the name. Doki provides auto-suggestions as the user types, allowing quick selection from a list of existing definitions (\autoref{fig:shot}).

References are linked to their source definitions. Any change to a definition automatically propagates through the document. For example, if \mention{@Pandi} is renamed to \mention{@Panda}, every occurrence of that character updates instantly. Similarly, if the description of a definition is revised or a new visual is attached, all references will reflect the updated content.

\begin{figure}[t]
\centering
\includegraphics[width=\linewidth]{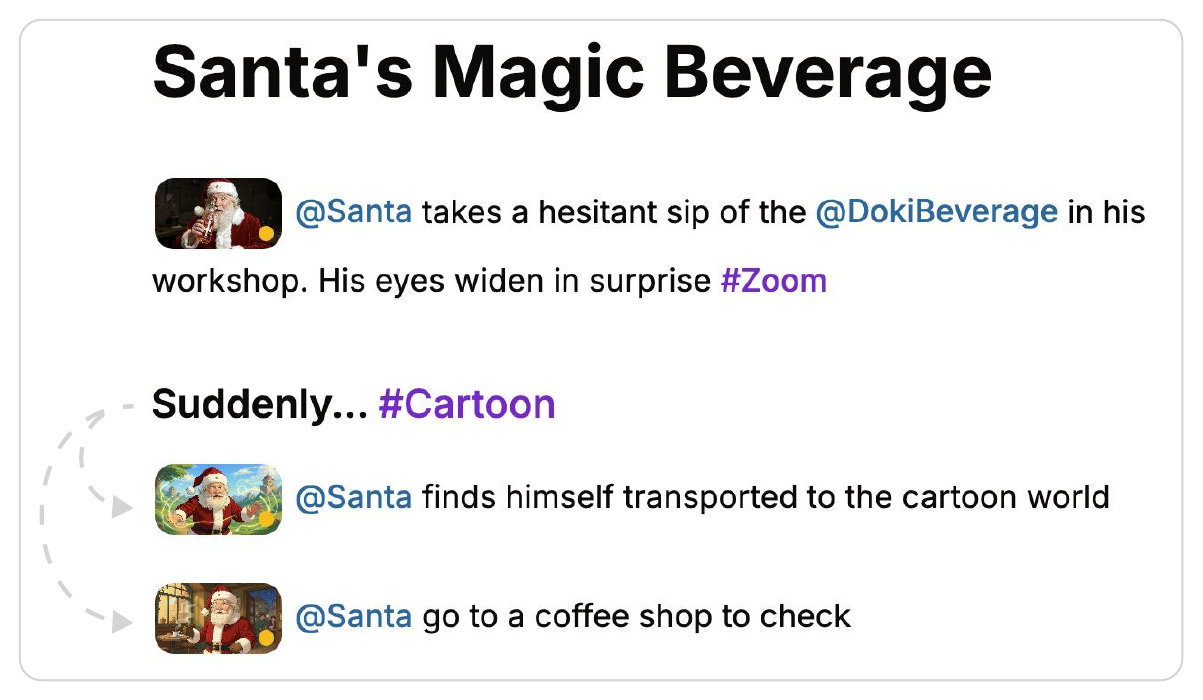}
\caption{Authors can apply definitions globally (\hashtag{\#all} \equalsign) or scope them to specific sections via headings. In the example, a scoped definition (\hashtag{\#Cartoon}) transforms subsequent shots under the heading level into a consistent cartoon style. }
\label{fig:headings}
\end{figure}
\subsubsection{Global and Scoped Definitions}
As documents grow in length in Doki, authors often need to apply the same definition, such as a \hashtag{\#Style} or \mention{@Scene}, across multiple paragraphs. Repeating these references manually can be tedious. To address this, Doki supports both \textit{global} and \textit{scoped} definitions that automatically propagate without requiring explicit referencing.

To define a global attribute, users assign the name ``all'' to any definition. For example, writing:
\begin{quote}
    {\inter \hashtagstylebadge\xspace \hashtag{\#all} \equalsign Photorealistic}
\end{quote}
automatically applies the photorealistic style to the entire document. This syntax works for all types of definitions, including mentions and hashtags.

In addition to global definitions, Doki supports heading-level scoping. Documents often contain a structured hierarchy of headings, and Doki leverages this structure to control the scope of definitions. Placing a definition on a heading applies that definition to all content within the corresponding heading level (\autoref{fig:headings}).
This design gives authors predictable control: for example, overall stylistic rules can be set once, and specific deviations can be introduced at finer levels without redundancy.

\subsubsection{Paragraph Handles}
\label{sec:handles}

When hover over a paragraph, up to two buttons may appear on the left. On the right, the handle \handleitem{\faGripVertical} allows users to reorder the paragraph or its associated shot sequence. 
The left-side handles communicate available generation options. If no shot is eligible for generation, the handle remains hidden. When the system detects that preview images can be produced, it displays a paintbrush icon \handleitem{\faPaintBrush}. If the paragraph is ready for video generation, the system instead shows a clapperboard icon \handleitem{\emoji{clapper-board}}.
Clicking the paintbrush triggers image generation for the paragraph. Clicking the clapperboard initiates video generation. The system does not regenerate all assets indiscriminately. Instead, it uses an algorithm to determine what needs to be generated and in what order, as explained in \autoref{algo:generateBtn} in Appendix.

\subsubsection{Audio}
\label{sec:audio}
Users can add audio to their shots through the slash command \slashmenu. From this menu, they may insert \audiospeech, \audiomusic, or \audiosfx. Authoring audio is via text in square brackets. The system interprets any text enclosed in brackets as audio. For clarity, the notation appears in gray and can be mixed with normal text.

When audio notations are included, they are passed to video generation models that support synchronized audio, such as Veo 3 and Veo 3 fast. If no audio is specified, the system produces no/minimal sound effects.

\begin{figure*}[t]
\centering
\includegraphics[width=\linewidth]{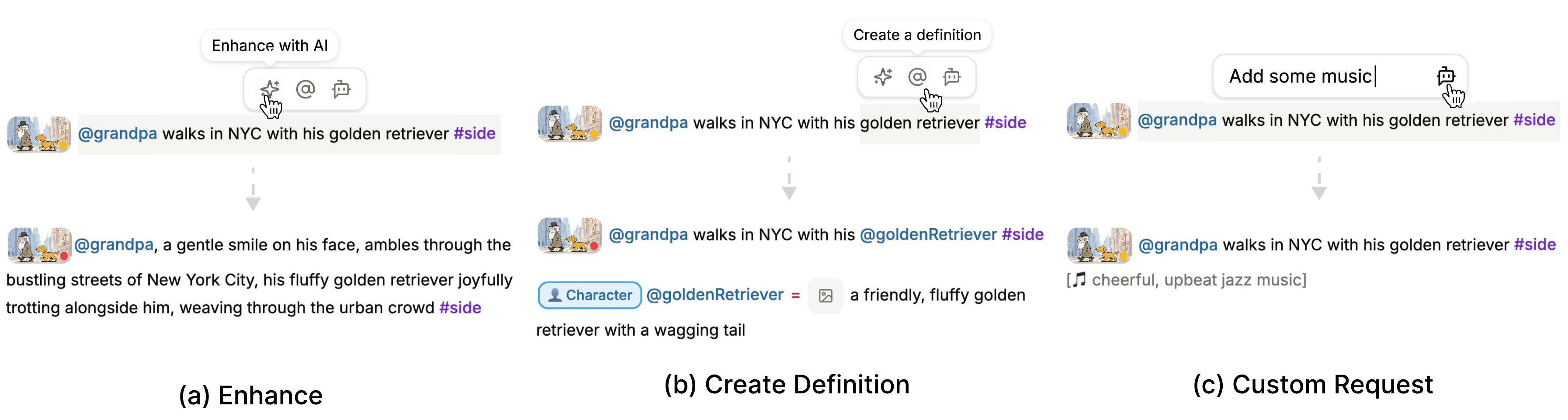}
\caption{Three types of inline agent actions. (a) ``Enhance'' increases the descriptiveness of the selected text. (b) ``Create Definition'' converts the selection into a reusable definition. (c) ``Custom Request'' applies an in-context edit based on user instruction.}
\label{fig:inline_agent}
\end{figure*}

\subsubsection{AI Agents}

Doki integrates two AI agents that provide complementary forms of assistance: the Sidebar Agent, which operates as a turn-based conversational assistant, and the Inline Agent, which supports direct, text-level edits within the editor. Both agents can perform small adjustments as well as large, multi-step edits. 
All edits are applied directly to the document and highlighted visually.
Implementation details of the two AI agents can be found in Appendix~\ref{appendix:agentic_edit}.

\paragraph{Sidebar Agent}
The Sidebar Agent can be opened from the main header and interacts with users through a conversational interface. Because the agent has full context of the document, it can carry out edits that span text, definitions, and generated media.

For example, it can support requests like:
\begin{itemize}
    \item Expanding concepts. \eg {\inter “Make the \hashtag{\#goldenHour} definition more detailed.”}
    \item Inserting visual references. \eg {\inter “Add a visual reference to the \mention{@castle} definition.”}
    \item Refining pacing and tone. \eg {\inter “Fix the pacing of the video. Make the first two shots more dramatic and add a climactic final shot.”}
\end{itemize}

\paragraph{Inline Agent}
The Inline Agent supports immediate, in-context editing within the document. It appears only when text is selected, showing a bubble menu with three options: Enhance, Create Definition, and Custom Request (\autoref{fig:inline_agent}): \textit{Enhance} increases the descriptiveness of the selected text. \textit{Create Definition} converts the selection into a reusable definition and replaces the text with that reference.\textit{ Custom Request} opens a small input field for user instructions.

With the inline agent, authors can begin with rough drafts or loosely organized ideas and progressively refine the drafts into structured narratives. 
For example, a writer might draft the sentence {\inter “A young boy runs across the courtyard.”} At first, the boy is an incidental figure, but the author later decides to establish him as a recurring character. By selecting {\inter “A young boy”} and clicking \icode{@}\xspace to turn it into a persistent \mention{@Character} definition. Then, the author can issue a custom request, such as {\inter “make him wear a hat”}, to expand the reference to add more details.

\subsubsection{Other Notable Features}
\begin{figure*}[t]
\centering
\includegraphics[width=\linewidth]{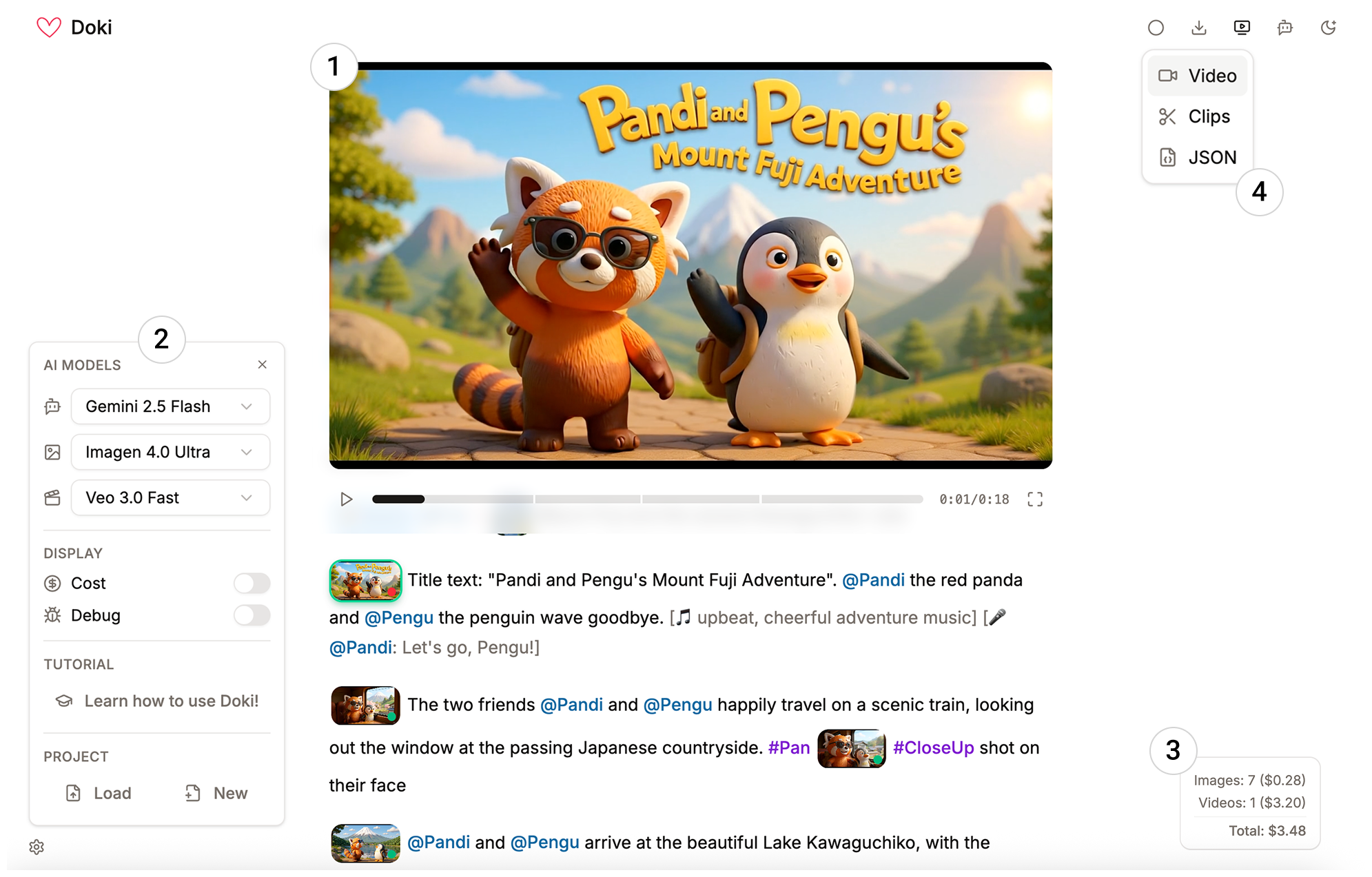}
\caption{Additional features in Doki: (1) Preview video player shows the generated video with shot segments.
(2) Settings panel lets users select AI models, toggle cost/debug options, access tutorials, and manage projects.
(3) Cost monitor tracks generation usage and expenses across images and videos in real time.
(4) Download menu provides output in video, clip, or JSON formats.}
\label{fig:utilites}
\end{figure*}

Doki also provides a set of other features to support the authoring process in the header and the settings panel (\autoref{fig:utilites}).


\noindent\textit{$\bullet$ Status Indicator}: A small circular icon communicates the current state of the system.

\noindent\textit{$\bullet$ Export}: Users can export their work at any time. The download menu supports three formats: the full rendered video, a zip file of all individual shots, and a JSON file containing the project's structured data. This enables seamless transfer across devices and external editing tools.

\noindent\textit{$\bullet$ Video Player}: A built-in video player allows users to preview their generated content without leaving the application. The player can be toggled open or closed and provides basic playback controls for reviewing the full sequence.
It also supports shot-level navigation.
As playback advances from one shot to the next, the system automatically scrolls the document to the corresponding section.

\noindent\textit{$\bullet$ Settings Panel}: The lower-left corner includes a settings menu for project-level actions. ``Load'' allows users to open a saved project from a JSON file. ``New'' starts a fresh project.  Users may select among supported AI models for text, image, and video generation. 




\begin{figure*}[t]
\centering
\includegraphics[width=\linewidth]{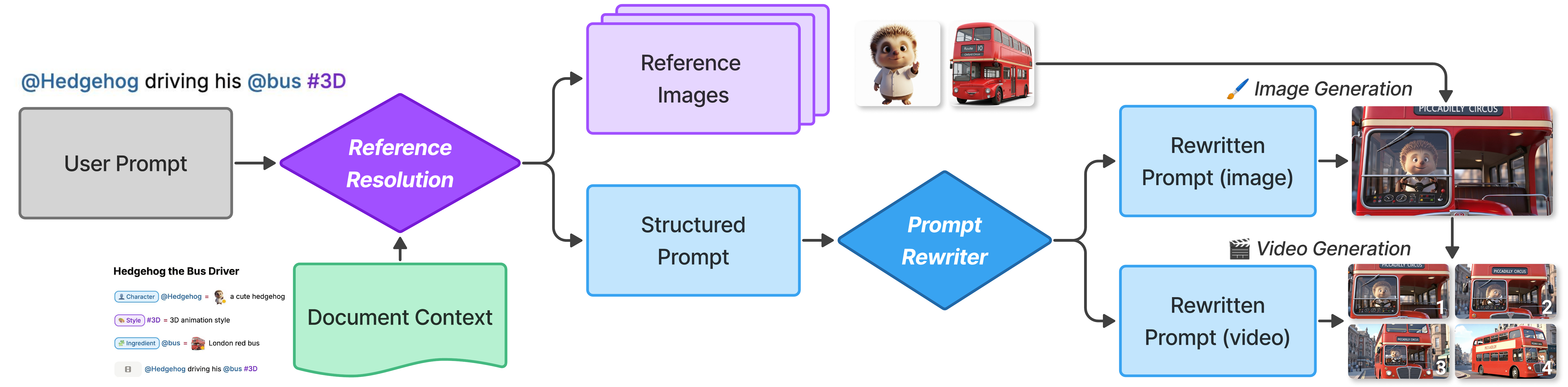}
\caption{Doki's shot generation pipeline. First, the user's raw prompt and the document context is passed to reference resolution module to create a structured prompt and gather relevant visual reference images from the definitions. Then, the prompt rewriter rewrites and polishes this structured prompt for image and video generation. Doki first produces a static preview image and then uses that image as a the first frame to generate the final video clip. }
\label{fig:pipeline}
\end{figure*}

\subsection{Doki's Shot Generation Pipeline}

To transform a user's text input into a video shot, Doki maintains three distinct types of prompts throughout this process~(\autoref{fig:pipeline}). 
The first is the \textit{user prompt}, which corresponds to the text entered directly into the editor. The second is the \textit{structured prompt}, which augments the user input with relevant references to definitions. The third is the \textit{rewritten prompt}, which refines and polishes the structured prompt to optimize performance for downstream image and video generation models.

\subsubsection{Structured Prompt}

Once the system receives the user prompt, it parses the text and resolves references to produce a structured prompt through a tree structure. This step grounds each reference in the prompt with its corresponding definition.
For example, for the user prompt:
{\inter \mention{@Hedgehog} driving his \mention{@bus} \hashtag{\#3D}},
the system transforms it into:

\begin{verbatim}
DESCRIPTION: @Hedgehog driving his @bus #3D
DEFINITIONS:
Character @Hedgehog = a cute hedgehog in a white shirt  
Ingredient @bus = London red bus  
Style #3D = 3D animation style  
\end{verbatim}

Instead of relying on the user prompt, Doki uses structured prompts to maintain consistency across revisions. For instance, updating the description of \mention{@Hedgehog} will automatically mark all shots involving that character as ``outdated'' (little red dot on bottom right), even if the user did not modify the shot text directly. 

\subsubsection{Rewritten Prompt}
\label{sec:prompt_rewriting}
Doki then refines the structured prompt through a prompt rewriter. 
The structured prompt is accurate but often rigid and unnatural. 
The rewriter addresses this by producing a fluent natural language description optimized for generative models.
In practice, the rewriter is implemented as a large language model prompted with carefully designed instructions. 
Prompts used for image and video rewriters are available in the supplementary materials.



\subsubsection{Selecting Image References}
Beyond resolving textual references and refining prompts, the system also retrieves relevant visual references to guide shot generation. 
When a definition includes an associated preview image, the system incorporates that image as a reference. 
In addition, when a paragraph contains multiple shots, the model retrieves the image from preceding shot in the same paragraph. This mechanism preserves visual continuity across a sequence, as illustrated in \autoref{fig:shot_sequence}.
In addition to rule-based strategies, the system uses a prompted LLM (full prompt in supplementary materials) to identify image references that are contextually relevant but not explicitly referred to. 
For instance, if a story introduces a location, shifts to another scene, and later returns to the original setting, Doki can infer that earlier images of the location should be referenced to maintain consistency. 

\subsubsection{Image and Video Generation}

Once the system produces a rewritten prompt and retrieves reference images, it can now generate images and video for a shot in Doki. 
As described earlier, generation always begins with a static image preview that serves as a first frame, and then from there generate the video.

For image frame generation, the model combines the rewritten prompt with the retrieved reference images as context. 
By default, Doki uses the \texttt{Flux Kontext Pro} model for image generation, which takes both text and images (up to 4) as input. 
Video generation defaults to \texttt{Veo 3 Fast}. 
Doki also supports a range of different models as listed in \autoref{appendix:models}.

\subsection{Implementation}
Doki is built with TypeScript and React for the frontend, TipTap for rich text editing,  Zustand for state management, Node.js for backend, and FFmpeg for video processing. 

\section{Diary Study}
\label{sec:deployment}
We conducted a week-long mixed-methods diary study of Doki. We designed our study around three research questions:

\begin{enumerate}
    \item[RQ1.] How do users perceive the benefits and limitations of Doki’s text-native interface? 
    \item[RQ2.] What workflows do people employ when using Doki? 
    \item[RQ3.] How does prior experience with video editing and generative AI influence the ways they perceive and use Doki?
\end{enumerate}

We chose a diary study over a lab study for three reasons. 
First, there is no established point of comparison for generative video authoring. Existing tools are designed for captured content and follow very different paradigms. 
Second, because generative video is new, most people have little prior experience, and a short session can only reveal initial impressions rather than deep insights. 
Third, Doki is a freeform system -- we expected participants to adapt the system and develop their own workflows, which can only emerge with extended use. 
A diary study offered the best conditions to help us learn these evolving practices.

\subsection{Participants}
We recruited ten participants (6 female, 4 male; Avg. age=32.4, SD=9.9, range=[24, 57]) for a week-long diary study of Doki.
Six were recruited from outside of our organization and four from within.
Roles included product designers (2), animators (2), a filmmaker, a graphic designer, a software engineer, a content creator, a UX designer, and a program manager (\autoref{tab:participants}).

Participants reported different levels of experience and frequencies of video creation: 4 created videos daily, 3 weekly, and 3 a few times per year. 
Their videos spanned diverse contexts, such as professional films, social media content (\eg TikTok, Reels, Shorts), academic/work projects, and personal vlogs.
Participants used multiple video editing tools. Adobe Premiere Pro was most common (8/10), followed by iMovie (5) and CapCut (5). Three reported using specialized generative video platforms (3/10, e.g., Runway, Pika, Google Flow).

Participants' experience with generative AI tools also varied. For large language models, 5 reported daily use while 3 shared rarely using them.
Image and video generation tools were less commonly used: 4 participants reported using image generation models at least once per week, and 2 reported using video generation models at least once per week. Notably, the participants who used video generation models regularly had professional roles closely tied to generative video creation: one is a filmmaker producing AI-driven video content; the other a product designer responsible for producing animation videos for their brand.
Overall, seven participants were already using generative AI in their existing video creation workflows. For example, scriptwriting with ChatGPT, storyboard ideation with Midjourney, refining edits with Photoshop's AI features, etc. 3 participants reported minimal or no prior experience with AI tools -- providing contrast in perspectives and practices.

\subsection{Procedure}

The study followed a three-phase structure: 

\paragraph{Phase 1: Onboarding (60 minutes)} 
Participants completed informed consent and a background questionnaire, received a guided tutorial to Doki's core concepts (e.g., shots, definitions, slash menu, agents) and completed a short open-ended creative task (``\textit{Create a 30-second video about anything}'') with support from one of the authors. 

\paragraph{Phase 2: In-the-Wild Diary Study (5 days)} 
Participants used Doki independently for five days. Each day, they were instructed to engage with the tool for at least 50 minutes, with the goal of producing 2-3 complete videos at the end of 5 days. Participants were encouraged to follow their natural workflow, allowing ideation, drafting, generation, and iteration to unfold across days. 

After each day, participants completed a brief survey.
To capture an overall measure of participant satisfaction, we included a single-item rating question (1-10): ``\textit{Please rate your overall satisfaction with your Doki experience today}''.
Participants also reported their workflow strategies, identified moments of ease and difficulty, and listed features used (\eg shot generation, definitions, AI agents, etc.). Each survey concluded with an optional submission of the day's project file and exported video.

In addition, we collected interaction logs spanning session duration; number of shots created or deleted; editing operations; words typed; use of mentions and hashtags, previews, and menus; generative activity (\eg attempted/successful image or video generations, regeneration counts, total generation time); export behavior; AI assistant interactions; and cost metrics based on image and video generations.

\paragraph{Phase 3: Exit Interview (60 minutes)} 
Participants completed a semi-structured remote interview. The first portion consisted of a retrospective think-aloud session~\cite{guan2006validity}, in which participants reviewed one of their projects, explaining creative decisions, feature use, and how the final video evolved. We also asked about overall experience, workflows, ownership, and learning. Participants also completed the System Usability Scale (SUS)~\cite{brooke1996sus} to provide standardized quantitative feedback.

\paragraph{Apparatus}
Participants accessed Doki remotely as a web app during the week-long study period. External participants were compensated \$30/hour. Collected data included submitted videos, Doki documents, surveys, interaction logs, and interview recordings/transcripts.

\section{Results}

\begin{figure*}[p]
\centering
\includegraphics[width=\linewidth]{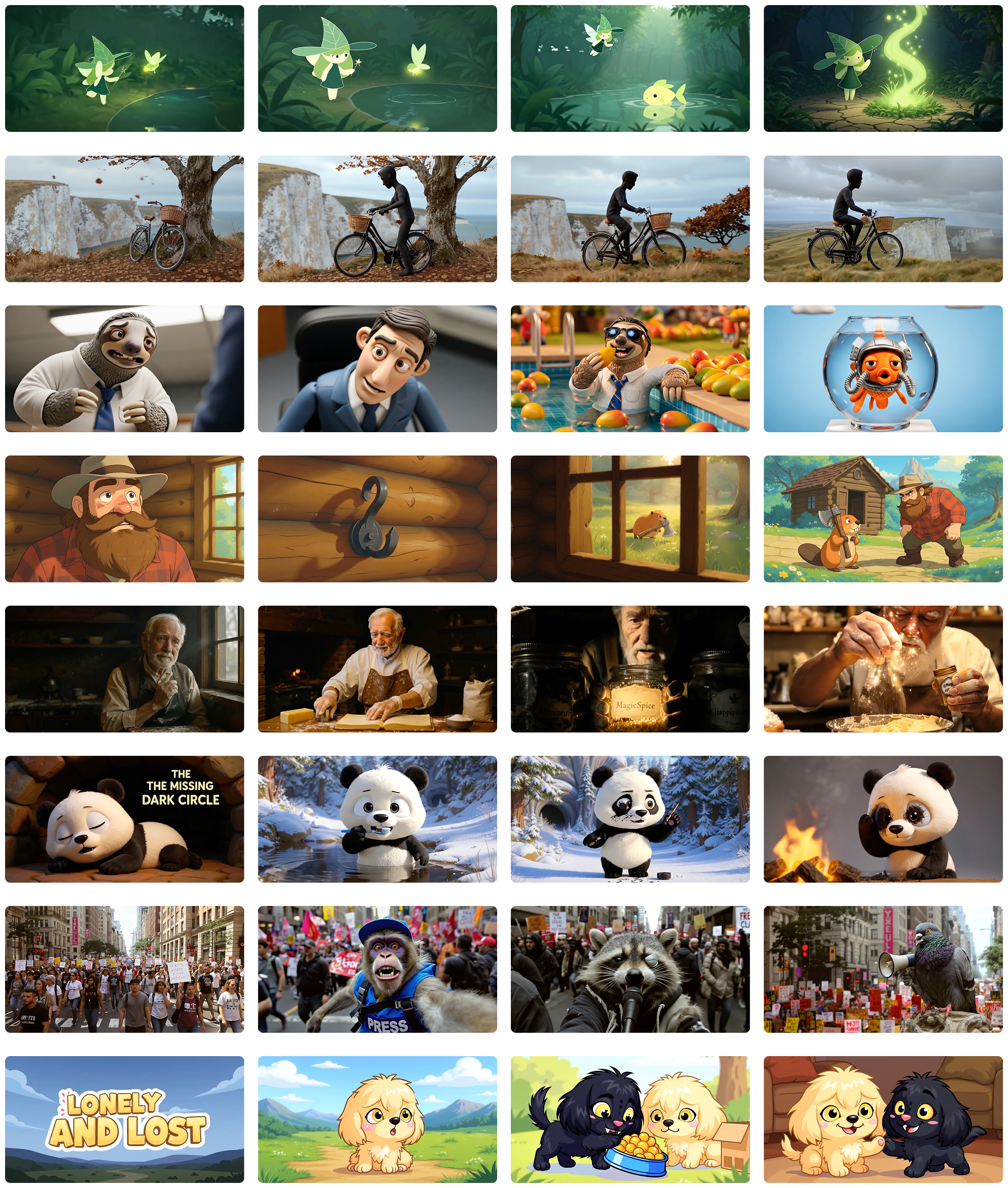}
\caption{Example frames from videos created by participants in our diary study. The videos demonstrate the breadth of video types Doki supports. Each row corresponds to a single video, shown left to right through four representative keyframes.}
\label{fig:videos}
\end{figure*}

\subsection{Doki Usage}

\subsubsection{Interface Usage}
\begin{figure*}[t]
\centering
\includegraphics[width=\linewidth]{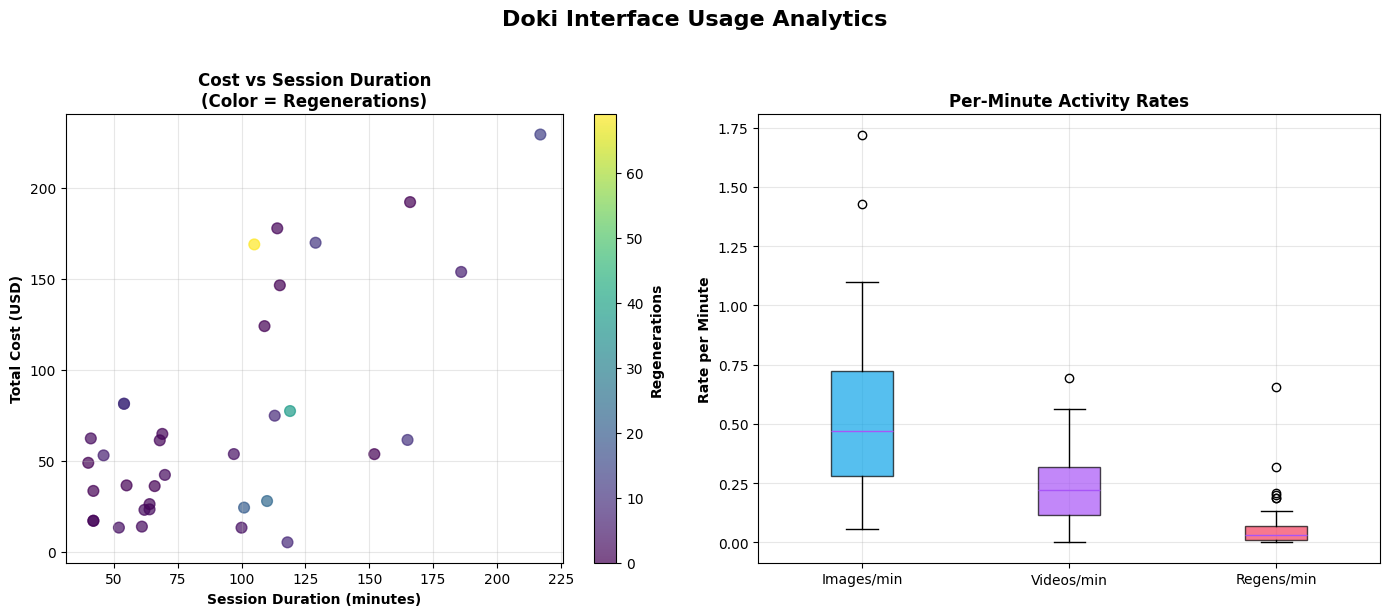}
\caption{Doki interface usage analytics. Left: Total cost by session duration, with point color indicating regenerations. Cost data are available for 35 of 50 sessions, as some sessions were logged under ongoing projects rather than new ones as instructed. Right: Distribution of per-minute activity rates for images, videos, and regenerations. Images show the highest median rate, followed by videos, with regenerations least frequent.}
\label{fig:usage}
\end{figure*}

Excluding the onboarding and exit interviews, participants spent an average of 91.7 minutes per session (SD = 45.5; range: 40–217) editing with Doki, indicating engagement beyond the required 1 hour per day. 
On average, they generated 45.5 images (SD = 33.2) and 20.3 videos per session (SD = 15.8). 
Normalized by time, participants produced approximately 0.6 images and 0.3 videos per minute (\autoref{fig:usage}). 

The most frequently used features were export (96\% of sessions), shot generation (90\%), and the video player (90\%). Participants also made extensive use of defining hashtags (82\%) and mentions (78\%). 
Shot variants were generated in 76\% of sessions, and audio was added in 68\%.
The chat sidebar AI agent was used in 74\% of the sessions while the inline agent was used 50\%. 
Heading-based styles were the least used feature, appearing in only 24\% of sessions.

\subsubsection{What Videos Did Participants Create with Doki?}
Participants submitted 46 video files in total. 
The average duration was 67.14 seconds (SD = 35.97; range: 15.04–184.04). Most videos (31) fell within 30-90 seconds; Four were shorter than 30 seconds, and eleven exceeded 90 seconds.

Participants produced a variety of videos (\autoref{fig:videos}) that fell into five main categories: storytelling, instructional video, advertising, music, and experimental. Storytelling was the most common, with eight participants (P2, P3, P4, P5, P7, P8, P9, P10) creating original narratives such as fairy tales, pet stories, or character-driven animations. One participant (P1) focused on instructional formats, producing cooking videos and academic explainers. Advertising and promotional content appeared in three cases: P2 created two claymation-style advertisement for Doki, P6 worked on a summer camp promotion video, and P8 produced a series of animal welfare videos tied to World Dog Day. 
Some participants explored less conventional directions. P8 experimented with music videos.
Two participants also created experimental projects like a surreal animal protest (P2) and ASMR (P7). 

While many participants used Doki for narrative storytelling, these projects demonstrate the breadth of video types Doki can support.
We include selected videos authored by participants in the supplementary material. 

\begin{table*}[t]
\resizebox{\textwidth}{!}{%
\begin{tabular}{@{}rlllllrrr@{}}
\toprule
\multicolumn{1}{l}{\textbf{Id}} & \textbf{Role} & \textbf{Video Creation Freq.} & \textbf{GenAI Use (Text)} & \textbf{GenAI Use (Image)} & \textbf{GenAI Use (Video)} & \textbf{SUS} \\ \midrule
1 & UX Designer & A few times a year & A few times a year & Monthly & A few times a year & 87.5 \\
2 & Filmmaker & Daily & Daily & Daily & Daily & 77.5 \\
3 & Product Designer & Weekly & Rarely or never & Rarely or never & Rarely or never & 75.0 \\
4 & Software Engineer & A few times a year & Daily & Weekly & Monthly & 90.0 \\
5 & Animator & Daily & Rarely or never & Rarely or never & Rarely or never & 80.0 \\
6 & Program Manager & A few times a year & Daily & Monthly & A few times a year & 72.5 \\
7 & Animator & Daily & Weekly & A few times a year & Rarely or never & 62.5 \\
8 & Product Designer & Daily & Daily & Weekly & Weekly & 87.5 \\
9 & Graphic Designer & Weekly & Monthly & Weekly & A few times a year & 90.0 \\
10 & Content Creator & Weekly & Daily & A few times a year & Rarely or never & 87.5 \\ \bottomrule
\end{tabular}
}%
\caption{System Usability Scale (SUS) scores and subscales (Usability and Learnability) across participant roles, video creation frequency, and generative AI tool use.}
\label{tab:sus}
\end{table*}

\subsection{RQ1. How do users perceive the benefits and limitations of Doki's text-native interface?}

\subsubsection{Overall impression of Doki}
Overall, participants reported a positive experience with Doki. The System Usability Scale (SUS) ratings (\autoref{tab:sus}) collected during the exit interviews averaged 81.2 (SD = 8.9, median = 83.8). This score corresponds to the \textit{Excellent} category defined by Bangor et al.~\cite{bangor2008empirical}, placing Doki within the 90–95th percentile (benchmark average = 68). Participant ratings ranged from a 62.5 to 90.0, with all but one participant rating Doki at or above 72.5.
These findings are consistent with daily survey measures of overall satisfaction, which had a mean of 7.48/10 (SD = 1.68, median = 8.0).

Beyond the quantitative measures, many participants described Doki as a robust system that generates high-quality visuals and is delightful to use. For example, P8 shared that ``\textit{This whole process makes me feel relaxed and gives me a great sense of accomplishment.}'' Similarly, P1 remarked, ``\textit{Doki is a really great generative AI video tool, probably one of the best I've used so far.}'' 

\subsubsection{Ease of use and learnability.}

Nine out of ten participants described Doki as very easy to use, highlighting its simple interface. They noted that the system required little effort to learn and that they could quickly grasp its core functionality. For example, P3 and P5 reported that common features such as the slash command, mention system, and hashtags mirrored those in platforms they already used daily. 
This familiarity helped them learn quickly and made the features easy to remember.
P8 found Doki much simpler than her usual video creation workflow because it minimized context switches: 
\begin{quote}
    ``\textit{Doki combines them together so I don't need to shift between different windows...it really saves me the window shifting time.}''
\end{quote}

These qualitative findings align with the survey results. Quantitative analysis showed no significant change in participants' satisfaction scores across the study period (Repeated Measures ANOVA, p = 0.42). Participants consistently rated their experience positively from the start. As P4 noted, ``\textit{Almost immediately… I felt like I could use it right away.}'' In the exit interviews, the System Usability Scale (SUS) subscale for learnability~\cite{lewis2009twofactor} yielded a median score of 81.2, which falls within the excellent range.

Because of Doki's simple interface, three participants mentioned that Doki made them \textit{focus on storytelling}. Compared to tools like Flash or After Effects, P9 described Doki as ``\textit{10 times easier},'' and he could focus on the idea and content rather than symbol animation or movement. As P5 put it, ``\textit{I think the interface is perfect...the whole interface is very simple and clean. So I can focused on creating my story.}''

\subsubsection{Fast to go from idea to content.}

Eight participants emphasized how Doki accelerated the process of moving from a rough idea to deliverable content. It enables participants to bypass many of the slow, manual steps in traditional workflows.
P1, a former professional documentary filmmaker, contrasted Doki with traditional production workflows: 

\begin{quote}
    ``\textit{For like a quick project, I think with my previous workflow that's just not possible because I have to go out to shoot, I have to spend a lot of time on editing. But with Doki I can just easily prompt the system to generate a very great first draft.}'' 
\end{quote}

Several participants mentioned that this efficiency was especially valuable for fast, lightweight projects where perfect visual fidelity was not the priority. P5 reported: ``\textit{So on average it will be 15 minutes to create about one minute or longer video.}'' Similarly, during the diary study, P8 produced a series of five 30-second videos for World Dog Day, in roughly ten minutes each -- an output that would have been unthinkable in their prior workflow.
Participants consistently noted that integrating an AI agent within the document interface was essential to this speed.
As P2 described, ``\textit{It can create short level breakdowns. It can auto format the whole scenario. It can do anything for you.}''

\subsubsection{Document representation fosters comprehension}
Four participants explained how Doki's document-based representation not only accelerated content creation but also enhanced their comprehension of narrative flow and structure. Unlike traditional non-linear editors (NLEs), which primarily rely on timelines, Doki offered a holistic, text-based overview of their projects. This shift was seen as particularly valuable in helping participants visualize their films in a semantic way. 
P2, a professional filmmaker, highlighted how this representation provided an immediate sense of coherence across the film:  
\begin{quote}
    ``\textit{Doki sets itself apart in how it gives you a sense of how your film will look. You can see the flow of your scenes, how your film is looking scene by scene and shot by shot.}''    
\end{quote}

By contrast, P3 noted that conventional NLEs often lack such an outline, making it harder to conceptualize story development beyond technical editing:  ``\textit{It (NLEs) doesn’t have a nice visual outline for writing or editing your videos or cropping.}''
Similarly, P6 explained that the document structure enhanced her understanding of video composition:
``\textit{I feel like it was helpful for me to understand the structure of [the] doc. It’s pretty intuitive.}''

In addition, participants mentioned that the text representation also made it very easy for them to view and understand AI's changes, and even for them to learn how to better use Doki (P4):
\begin{quote}
    ``\textit{The notebook interface with text as the driving element is clever. You can learn from how it generates music or voice pieces, even if you wouldn't use it directly.}''
\end{quote}

\subsubsection{Parametrization makes story building easier}

Six participants emphasized how Doki’s parametrization design helped them build their stories. 
For example, by defining reusable elements such as characters, scenes, and styles, they could begin projects with a strong foundation even when the narrative was not yet fully formed. As P6 explained, ``\textit{Building reusable characters and ingredients really made sense as a lot of times I have some clarity into what/who I want to use but not the story itself.}''

Participants also described how these definitions supported coherence across their work. P10 highlighted that parametrization reduced randomness in generative outputs. 
``\textit{I think this tool is really cool because you can define a character and scene so it would not be like very randomly generating the video.}''
Similarly, P1 noted that Doki itself is not just a video tool but also a really good tool for prompt engineering and crafting the visuals, even just a single shot. 
He described that the definition system make it easier for him to build up complex scenes from smaller components: 
\begin{quote}
    ``\textit{I don’t have to write repeated keywords just to keep this consistent. I can just use tags and add each character so and they will stay.}'' 
\end{quote}
He further envisioned that in the future, professional creators could design and share parametrized styles or camera movements as creative assets, and make them available through simple tags for others to use. 

Finally, participants appreciated how parametrization integrated naturally into Doki's text-native representation. P5 noted how features like mention, hashtag and audio functioned seamlessly within the document: ``\textit{...mention and hashtag and create audio. It's all in the same logic thing. It’s very easy to remember and learn.}''


\subsubsection{Precise control is limited.} Nine participants noted that Doki’s outputs often failed to match their prompts, requiring repeated regeneration to achieve acceptable results. 
P1 described persistent artifacts: ``\textit{A lot of videos include nonsensical text that I couldn’t remove, even when I explicitly prompted it.}'' Such errors forced participants into trial-and-error workflows. As P10 explained, ``\textit{I need to regenerate image again and again...
}'' P8 also noted visual glitches such as ``\textit{dogs disappearing mid-run and people brushing the air instead of the animals.}''
Several expressed frustration that the text-only interface constrained their ability to realize specific visual goals. P9 noted, ``\textit{Doki is not good at when you want specific frame compositions.}'' This challenge was especially pronounced for users with \textit{strong visual references}. During the study, P10 attempted to create a video from her storybook. Because she already had a clear mental image of the visuals and compositions, she found it difficult to reproduce them through text prompts. She explained, ``\textit{If you want to create your personal story, sometimes the video could not reflect everything that you want. Maybe just achieve 80\% of your imagination.}'' 

\subsubsection{Hard to work with audio and music.} Participants noted constraints in how Doki handles time-based media. Because the system is organized around a paragraph and document structure, it is difficult to add audio that spans multiple paragraphs or begins asynchronously. P8, who attempted a music video, found it hard to align visuals with audio.

\subsection{RQ2. What workflows do people employ when using Doki? }

\subsubsection{A lot of variety in how people start.}

Participants entered Doki from different starting points, reflecting varied creative practices. Some began with a complete script that they had written beforehand. For example, P1 noted: 
``\textit{I already have a text version of that recipe, so I just use the chatbot, paste the whole recipe and ask [it] to generate all the ingredients I may need.}''

Others would first defining characters, scenes, or other assets before turning to narrative construction. As P3 explained:

\begin{quote}
``\textit{I’d probably define all the assets and the scenes. Then once I did that, I would tell Doki the story I want to tell and have it fill in the rest, like the story outline for me.}''
\end{quote}

A third group described a more exploratory and freeform approach. For these participants, Doki supported writing and discovery without preparation. P4 mentioned that he can ``\textit{just begin writing something.}'' Similarly, P2 worked in small pieces and refined step by step:

\begin{quote}
``\textit{I worked on small pieces iteratively, building scenes and refining character interactions step by step.}''
\end{quote}

Across these practices, Doki enabled participants to begin video authoring in ways that matched their own practices. Whether starting from a script, from defined assets, or from improvisational writing, participants found that the system supported flexible entry points into the creative process.

\subsubsection{High reliance on AI}
Eight participants reported that they would begin by using AI to generate a first textual draft before refining further. 
Six participants went further, relying almost entirely on Doki's AI agents throughout the process. 
They described starting with a rough, often one-line idea, using the AI as a jump start, and then remaining in the conversation panel or inline agent to complete their projects with minimal manual edits.
P8 noted that she often preferred asking the AI to revise drafts rather than editing them manually. As she explained, ``\textit{AI knows better how to talk to AI.}''
P10 described a similar reliance on delegation: ``\textit{I don’t really read the script the agent creates. I just want to see how the image would look...I generate those images and then see if the agent misunderstood my description.}''

This pattern was not limited to casual creators. P2, a professional filmmaker who produced some of the best videos in our study, revealed that nearly the entire process was conducted with the AI: ``\textit{``I had just one line idea...and it really helped me to brainstorm much more in terms of how different scenes will be, how we can create different shots...and then we worked together to turn [it] into a full script.}''

Four participants described a gradual shift from manual work to heavier AI reliance as they grew more comfortable and found it reliable and easy to use. P4 explained: 
\begin{quote}
    ``\textit{The workflows are slightly different. At first, I did that very manually. The second one was more hybrid type, and then I'll move to AI more and more, for the last video, I’ll just type the command, tell AI what to do rather than manually edit.}''
\end{quote}

We interpret this increasing reliance not only as a reflection of participants' trust in the system, but also as a side effect of the design of Doki itself. By making collaboration with AI extremely simple through a text representation, participants can find it natural to delegate more and more of the process to AI. 
In other words, simplicity of interaction could encourage reliance.

\subsubsection{Users who rely almost entirely on AI still feel ownership.}
An interesting finding is that, using Doki, even participants who relied almost completely on AI still felt \textit{strong sense of ownership} of their videos. P4 cited Andy Warhol’s saying:
\begin{quote}
    ``\textit{Art is the process of an artist selecting}.''
\end{quote}
P1 similarly explained: ``\textit{Because I get to decide what every item looks like, I feel like this is my creation. That’s what gives me ownership.}''
Participants often compared themselves to directors, as P6 described that she feels her role was no different from a director's, except working with an AI agent instead of human agents.

Several participants also contrasted this experience with other platforms. P10 noted, ``\textit{When I use MidJourney, if other people use similar prompts, they would get 90\% similar images. I would think this is generated by MidJourney, so I give the credit to MidJourney. With Doki, I feel ownership. Every choice of image or cut made me feel I’m a video maker.}''
Participants found that Doki enabled a workflow where creators could delegate most production to AI while still experiencing a strong sense of authorship. 

Two participants did not use AI at all. Both were animators who typically worked by hand and had little prior experience with text-to-image or text-to-video models. They expressed a preference for full creative control: ``\textit{I just want to create the story by myself.}'' (P5). 

\subsection{RQ3. How do users' prior experience with video editing and generative AI influence the ways they perceive and use Doki?}

\subsubsection{Doki empowers participants without previous experience to create video stories}
For participants without training in filmmaking, animation, or generative AI, Doki opened up new possibilities that had previously felt out of reach. Four participants noted that they would never have created a video story without Doki, and now, they could express ideas in new ways. For example, P10, who had no prior drawing or animation experience, described:
\begin{quote}
    ``\textit{Like just use this and generate some animation. I do not have experience drawing those, but I can use this tool to create that.}''
\end{quote}

Others emphasized Doki's role in unlocking creativity and enabling personal expression. P9 called it ``\textit{my dream tool},'' explaining that it helped deliver messages and bring childhood stories to life. 
For many participants, Doki meant more than just a tool -- it represented a shift in what creative expression could mean for them. As P1 reflected:
\begin{quote}
``\textit{It kind of [offers] a new capability I didn’t have before. In the past, if I want to share something with my friends, I have to write it down, but now I can create a video to express whatever I want.}''
\end{quote}

\subsubsection{Experts position Doki as complementary rather than substitutive}

Participants with professional experience in video editing described Doki as fundamentally different from their existing tools. Rather than viewing it as a replacement, they see it as complementary for projects outside their primary workflows.
P1 explained, 
\begin{quote}
    ``\textit{I think creating a video with Doki is very different than my previous video editing experience. I would even define them as two completely different tasks, and are for very different purposes.}''
\end{quote}

Others emphasized this difference by situating Doki within the production pipeline. P4 viewed traditional editing tools to be essential for precision and polish, while Doki was valued for its speed in ideation and draft generation.
Similarly P5 stated:
\begin{quote}
    ``\textit{I think I will use [Doki in] the first period of my animation making process because I can first have a script and then input it into Doki to create a video for me to as a reference.}''
\end{quote}
Yet even as professionals stressed that Doki could not substitute for high-fidelity outputs, they also highlighted the dramatic contrast in efficiency. P5, who normally works through labor-intensive frame-by-frame animation, reflected that producing one minute of video by hand could take two months, while with Doki she could create it in about an hour. For her, the speed advantage was undeniable, even though the resulting output still carried a ``\textit{clearly AI feeling.}''

Another recurring challenge for professionals was that their existing expertise felt difficult to transfer into Doki. P10 reported, ``\textit{Although I learned some filming technique, I cannot use it here. Even if I have some idea in my mind, I don’t know how to describe in words.}'' Similarly, P7 expressed frustration that drawing skills could not be applied directly.
For professional with high quality standards, the visual quality Doki can generate still falls short:

\begin{quote}
    ``Animation industry standard is way higher for now...for animators, this is far from enough.'' -- P7
\end{quote}

\subsubsection{Prior experience with generative tools leads to greater satisfaction}

Participants with previous experience in generative AI video tools reported greater appreciation for Doki's capabilities. P1, who had recently worked on a generative video research project, explained:
\begin{quote}
``\textit{I tried a lot of different tools, including Google’s Veo 3, Google Flow, and a tool that we developed, [but] the consistency that I can generate across the scenes is just uncanny. Doki did something that was not previously possible.}''
\end{quote}

Similarly, P2 noted that long-term familiarity with generative tools gave him a clearer sense of their strengths and limitations, a perspective that new users may lack:
\begin{quote}
``\textit{Now that I have been using these tools for more than two years, I still don’t have an idea how good they are with each of these styles...new users would have no idea.}''
\end{quote}

Film knowledge also emerged as an important factor. Participants with professional or semi-professional backgrounds were more adept at structuring stories and translating them into coherent visual sequences. As P2 explained,
\begin{quote}
``\textit{It usually requires a craftsmanship which most people don't have. People do have ideas, but people don’t know how to create an entire piece. That’s where expertise comes in.}''
\end{quote}

\section{Discussion and Future Work}

While text-native authoring made it easier for participants to get from idea to video, we found that a low-barrier to creation does not on its own result in high-quality storytelling.  



\paragraph{Approachability does not guarantee quality.}
A text-native interface lowers barriers, but strong stories still demand craft. In our diary study, participants moved quickly from concepts to completed videos, yet not all videos had compelling arcs. 
Doki’s document-centric structure makes working with narrative elements like characters and scenes explicit and editable. 
Expert film makers know how to structure a narrative and how to sequence shots and scenes, while novices are left unsure how to improve their video. 
Future work includes optional narrative scaffolds (\eg three-act templates, A/B/C plots) and lightweight diagnostics (pacing irregularities, under-specified protagonists) that help authors see and improve a story without adding UI complexity.


\paragraph{Do we really need a video model for longer videos?}
Current text-to-video systems cap out at short durations (5–10 seconds\cite{veo3,runway}). 
At first glance, this seems like the primary limitation. 
Yet the average shot duration in contemporary film is roughly 4 seconds~\cite{cutting2016evolution}, well within today model's capability. 
Our formative and diary studies suggest that the main gaps lie elsewhere: \emph{consistency} (stable characters, props, styles), \emph{control} (camera grammar, timing, transitions), and \emph{context} (memory across shots and sequences). 
In Doki, we treat duration as a compositional rather than monolithic model problem. 
Text generates \emph{keyframes} that anchor identity, layout, and style; these keyframes expand into short, controllable shots. 
Longer models will still matter for specific cases (e.g., “oners,” dance, sports, talking heads), but for many scenarios, progress on cross-shot memory, identity preservation, and story-aware control is likely to deliver more value than raw clip length.


\paragraph{Document as a human–AI common ground.}
Most current AI creation tools such as Lovable (text-to-website), Deep Research (text-to-report), or Eleven Labs (text-to-audio), follow a direct input–output model: a prompt produces a finished artifact with limited opportunities for inspection or revision along the way. By contrast, Doki introduces the document as an intermediate representation: a format that is simultaneously readable and editable by humans and interpretable and executable by AI.

In future iterations, even if Doki begins from a simple prompt box, the output would not be a locked timeline or rendered video, but rather a document representation: a shared working space between human and AI where operations remain legible, revisable, and easy for user intervention.

Doki's document representation approach highlights a broader opportunity for the HCI community: how to design effective \textit{intermediate spaces} that make AI activity transparent and at the same time create richer opportunities for human agency.

\paragraph{Temporal expressivity limits of a document representation.}
\label{sec:concurrency}
A linear document is limited in expressing any temporal concurrency, cross-cut constructs, and duration-sensitive rhythm. Participants struggled to specify overlapping or offset audio (e.g., pre-lapped dialogue, music that bridges sequences), and transitions that bind across shot boundaries (e.g., J/L cuts, cross-dissolves, match cuts) without resorting to external tools. We implemented inline audio notations (\S\ref{sec:audio}) to address basic needs, but finer control remains awkward. One path forward is to extend the Doki language with lightweight temporal primitives that preserve readability while increasing control: inline event offsets (e.g., {\inter [SFX door slam \@ +1.2s]} or {\inter [MUSIC fade in 4s]}), paragraph- or sequence-scoped beat/tempo markers (e.g., {\inter \hashtag{\#Tempo}=110 BPM}), and parameterized transition definitions that bind explicitly to adjacent shots (e.g., {\inter \hashtag{\#CrossDissolve} 12f}, {\inter \hashtag{\#JCut}}). 



\paragraph{Compositional structures vs. single representation}
Most video authoring tools rely on ``compositional structures'' with interconnected underlying data. Professional non-linear editors such as Premiere Pro and Final Cut present projects as a network of linked views: an asset library, a multi-track timeline, a source monitor, and often a transcript panel for editing by text. 
Descript~\cite{descript} aligns transcript editing with a timeline. 
Recent research systems adopt the same principle.
VideOrigami~\cite{cao2025compositional} links scripts, storyboards, canvases, and timelines.

Doki explores a radically different direction. Instead of coordinating across multiple views, it centralizes all representations within a single text-native substrate. 
Authoring video is reframed as writing. 
This presents a different set of tradeoffs.
Multiple views can indeed support task-specific efficiency, such as precise pacing control in a timeline or spatial layout on a canvas. 
But Doki, for a lot of people whether they are professionals or amateurs, greatly simplifies the interface and interaction: there are very few mechanisms to learn and people can begin authoring almost immediately. 
This unified interface also strengthens version control and collaboration between humans or AI, while introducing challenges in temporal limits and etc.
From an HCI perspective, these paradigms are not in opposition but explore distinct possibilities. 
In a sense, they embody two different philosophies: compositional structures seek to optimize every step of authoring with specialized views, while Doki leverages AI to pursue an ultimately minimalistic interface.
\section{Conclusion}

In this paper, we introduced Doki, a text-native interface for authoring generative videos.
Findings from a week-long diary study show that it enables faster idea-to-content workflows, improved coherence via parameterization, and clearer narrative comprehension compared to traditional tools.

More broadly, we believe that Doki contributes a paradigm that positions text not only as input to generative systems but as the primary substrate for narrative, structure, and production.
We believe that as generative models continue to expand in capability, it is essential to rethink interface paradigms for content creation. 
Doki demonstrates one such direction -- a low-barrier, text-native model of authoring -- and opens new questions about how natural language might serve as the foundation for future creative tools.

\bibliographystyle{ACM-Reference-Format}
\bibliography{refs}

\section{Appendix}

\subsection{Technical Details}
\subsubsection{JSON example}
\label{appendix:json}
We show a simplified example JSON file of Doki to illustrate its underlying data structure:
\begin{lstlisting}[language=json]
{
  "shots": [
    {
      "id": "shot-1757235383722-pynguler9",
      "context": "definition",
      "structuredPrompt": "CHARACTER: corgi, a cute corgi with golden-brown and white fur.",
      "status": "video-ready"
    },
    {
      "id": "shot-1757234503129-4esljq3y6",
      "context": "definition",
      "structuredPrompt": "LOCATION: airport, a small regional airport terminal.",
      "status": "image-ready"
    },
    {
      "id": "shot-1757235144859-7k10hdcyn",
      "context": "paragraph",
      "structuredPrompt": "DESCRIPTION: corgi arrives at the airport with luggage.",
      "status": "image-ready"
    },
    {
      "id": "shot-1757234758231-ldfx5tr7d",
      "context": "paragraph",
      "structuredPrompt": "DESCRIPTION: corgi then boards the plane.",
      "status": "image-ready"
    }
  ],
  "tags": {
    "mentions": [
      { "name": "corgi", "tagName": "Character" },
      { "name": "airport", "tagName": "Scene" }
    ],
    "hashtags": [
      { "name": "all", "tagName": "Style", "userPrompt": "Japanese anime style" }
    ]
  }
}
\end{lstlisting}

\subsection{Agentic Editing}
\label{appendix:agentic_edit}
Doki supports agentic editing that allows both the sidebar and inline agents to directly manipulate the document. 
This enables authors to interactively revise content through natural language commands. 
Here, we detail the implementation of the JSON-based editing API.

\subsubsection{Document Context and Agent Memory}
Both agents in Doki have access to a structured representation of the full document. This context includes definitions, headings, paragraphs, shots, and associated metadata.
The sidebar agent also maintains a conversational memory that supports multi-turn interactions.

\subsubsection{Editing API}

Doki's AI agent editing feature is built around a structured API. All edits -- whether triggered through chat or inline interaction -- are represented as typed JSON objects. Each specifies:
\begin{itemize}
    \item \textit{id}: a unique identifier for the edit operation
    \item \textit{target}: the type and id for the document node to be modified (\eg definition, shot, paragraph, heading)
    \item \textit{newContent}: the replacement content (or null for deletions)
    \item \textit{description}: a concise explanation of the edit that will be shown to the user
\end{itemize}

This JSON format enables a wide range of editing operations such as updating definitions, revising shots, replacing paragraphs, modifying headings, inserting or deleting content, and performing a sequence of multiple edits. 

For example user asks, \textit{``Add an establishing shot right after the introduction.''}  
The agent translates this request into an insertion edit at the appropriate location. The JSON representation specifies the insertion point (after the introduction paragraph) and the new content to be added:
\begin{lstlisting}[language=json]
{
  "id": "edit-5",
  "target": { "type": "insert", "selector": { "after": "paragraph_intro" } },
  "newContent": "@Hero enters the castle gates #WideShot",
  "description": "Inserted a new establishing shot after the introduction paragraph"
}
\end{lstlisting}

Here, we show another example of how we implemented the Create Definition feature for inline agent (\autoref{fig:inline_agent}). When the user highlight the text \textit{``a dark sorcerer with glowing red eyes''} and click \icode{@}\xspace in the bubble menu,
Doki generates two coordinated edits and executes them in sequence. The first creates the new definition, and the second replaces the original text with a reference to that definition:

\begin{lstlisting}[language=json]
{
    {
      "id": "edit-6a",
      "target": { "type": "insert", "selector": { "after": "paragraph_12" } },
      "newContent": "Character @Villain = a dark sorcerer with glowing red eyes",
      "description": "Created a new character definition from selected text"
    },
    {
      "id": "edit-6b",
      "target": { "type": "paragraph", "selector": { "nodeId": "paragraph_12" } },
      "newContent": "The @Villain raises his staff to cast a spell",
      "description": "Replaced original text with reference to the new definition"
    }
}
\end{lstlisting}

\subsubsection{Shot Generation Logic}
\autoref{algo:generateBtn} shows a pseudo code of the logic of deciding what shot and what media type to generate for each paragraph, depending on the shot states. 
\begin{algorithm}
\caption{Shot Generation for Paragraph $P$}
\label{algo:generateBtn}
\KwIn{Paragraph $P$ with shots $s \in S$}
\KwOut{Generate button for $P$}
\BlankLine

\If{$\exists s \in S$ currently generating (image or video)}{
    \Return \faSpinner\xspace \texttt{Loading} icon\;
}
\ElseIf{$\exists s \in S$ can generate video}{
    \Return \emoji{clapper-board} \texttt{Clapper Board} button 
           (click to generate video for eligible shots)\;
}
\ElseIf{$\exists s \in S$ can generate image}{
    \Return \faPaintBrush\xspace \texttt{Paint brush} button 
           (click to generate image for eligible shots)\;
}
\Else{
    \Return no button\;
}
\BlankLine
\textbf{Definitions:}\\
\Indp
\textit{A shot can generate video} if it is not a definition shot, 
has an image ready (or outdated, or a saved image), 
and is not already generating.\\
\textit{A shot can generate image} if its status is idle, ready, outdated, or error.
\end{algorithm}

\subsection{Supported Models}
\label{appendix:models}
Doki by default uses Gemini 2.5 Flash for text, Flux Kontext Pro for images, and Veo 3 Fast for videos, while also supporting a broad range of generative models. 
For text-based interactions (image and video rewriting, AI agents), Doki supports Gemini 2.5 Flash and Pro. 
Image generation models include Imagen 4.0 Fast and Ultra, Ideogram 3.0, Flux Kontext Pro and Max, and Gemini 2.5 Flash Preview (nano banana). Among these, Flux Kontext Pro, Flux Kontext Max, and Gemini 2.5 Flash Preview accept images as context, while the remaining models operate solely on text. 
Video generation is enabled through Veo 2.0, Veo 3.0, Veo 3.0 Fast, and Runway Gen4. Veo 3.0 and Veo 3.0 Fast can produce synchronized audio.
Veo models generate 8-second clips and Runway Gen4 generates 5-second clips.

\subsection{Prompts}
\label{appendix:prompts}
\subsubsection{Image Rewriter:}
This module transforms structured prompts into vivid, detailed descriptions suitable for image generation. 
It takes in structured prompt and document context, and rewrites the prompt into expressive natural language while preserving fidelity to the source definitions. 
The rewriter ensures narrative consistency based on document context and adapts the description across diverse content types, including character appearance (\eg will use plain background), settings, visual style, and camera shots. 

\subsubsection{Video Rewriter:}
The video rewriter performs an analogous role for video generation. It produces detailed, coherent prompts that expand the structured input into cinematic sequences. Unlike the image rewriter, which focuses on visual descriptiveness, the video rewriter emphasizes motion and temporal flow. 
Prompts specify subjects, actions, camera positions, compositions, and relevant audio when explicitly mentioned. 
We also leverage cinematography principles to guide shot design such as camera motion and composition. 
The video rewriter also includes negative prompts to prevent undesired elements such as background music, speech, or text overlays if not specified by users.

\begin{table*}[t]
\resizebox{\textwidth}{!}{%
\begin{tabular}{@{}rrllllllll@{}}
\toprule
\multicolumn{1}{l}{\textbf{Id}} & \multicolumn{1}{l}{\textbf{Age}} & \textbf{Gender} & \textbf{Role} & \textbf{Video Types} & \textbf{Video Editor Use} \\ \midrule
1 & 28 & M & UX Designer & Professional films & DaVinci Resolve, GenAI tools \\
2 & 32 & M & Filmmaker & \begin{tabular}[c]{@{}l@{}}Social media, \\ Professional films\end{tabular} & Adobe Premiere Pro, Final Cut Pro \\
3 & 24 & F & Product Designer & \begin{tabular}[c]{@{}l@{}}Social media, \\ Personal story or vlogs\end{tabular} & Adobe Premiere Pro, iMovie, CapCut \\
4 & 57 & M & Software Engineer & \begin{tabular}[c]{@{}l@{}}Personal Story or vlogs, \\ Music video\end{tabular} & Adobe Premiere Pro \\
5 & 28 & F & Animator & \begin{tabular}[c]{@{}l@{}}School projects, \\ Work-related content, \\ Professional films\end{tabular} & Adobe Premiere Pro \\
6 & 37 & F & Program Manager & \begin{tabular}[c]{@{}l@{}}Social media, \\ Personal story or vlogs\end{tabular} & CapCut \\
7 & 24 & F & Animator & \begin{tabular}[c]{@{}l@{}}School projects, \\ Professional films\end{tabular} & Adobe Premiere Pro, iMovie \\
8 & 29 & F & Product Designer & \begin{tabular}[c]{@{}l@{}}Social media, Personal story or vlogs, \\ School projects, Work-related content\end{tabular} & \begin{tabular}[c]{@{}l@{}}Adobe Premiere Pro, Final Cut Pro, \\ iMovie, CapCut, GenAI tools\end{tabular} \\
9 & 38 & M & Graphic Designer & \begin{tabular}[c]{@{}l@{}}Social media, \\ Personal story or vlogs\end{tabular} & \begin{tabular}[c]{@{}l@{}}Adobe Premiere Pro, DaVinci Resolve, \\ iMovie, CapCut, GenAI tools\end{tabular} \\
10 & 27 & F & Content Creator & Social media & Adobe Premiere Pro, iMovie, CapCut \\ \bottomrule
\end{tabular}
}%
\caption{Participants information including demographics, roles, video practices, and generative AI tool usage.}
\label{tab:participants}
\end{table*}

\end{document}